\documentclass[sigconf]{acmart}

\AtBeginDocument{%
  \providecommand\BibTeX{{%
    \normalfont B\kern-0.5em{\scshape i\kern-0.25em b}\kern-0.8em\TeX}}}

\usepackage{longtable, acmart-taps}
\usepackage{booktabs,arydshln}
\usepackage{makecell}
\usepackage{soul} 
\usepackage{colortbl,booktabs}
\usepackage{fontawesome5}
\usepackage{array} 
\usepackage{siunitx}
\usepackage{wrapfig,lipsum,booktabs} 
\usepackage{tikz}
\usepackage{array} 
\usepackage{enumitem}
\usepackage{dblfloatfix}
\usepackage{csquotes} 
\usepackage{soul}
\usepackage[colorinlistoftodos,prependcaption,textsize=footnotesize,textwidth=2.6cm,bordercolor=orange,backgroundcolor=yellow,disable]{todonotes} %add 'disable' option to hide them all
\definecolor{myhighlight}{RGB}{242,242,242}
\definecolor{trustee_color}{HTML}{ED4F6F} 
\definecolor{state_color}{HTML}{51AFBC} %129F57
\definecolor{trustee_attributes_color}{HTML}{4384F0}

\definecolor{trustee_info}{HTML}{ED97A7}  %F1CBD2 
\definecolor{trustee_user}{HTML}{A4E1E9} %C7E4E8
\definecolor{trustee_platform}{HTML}{F6BE83} %F8E1C9
\definecolor{trustee_other}{HTML}{A4A4A4} %D5D5D5

\definecolor{highlight}{HTML}{fff44f}

\def\mybar#1{%%
  {\color{gray}\rule{#1cm}{6pt}}} 

\makeatletter
\def\adl@drawiv#1#2#3{%
        \hskip.5\tabcolsep
        \xleaders#3{#2.5\@tempdimb #1{1}#2.5\@tempdimb}%
                 #2\z@ plus1fil minus1fil\relax 
        \hskip.5\tabcolsep}
\newcommand{\cdashlinelr}[1]{%
  \noalign{\vskip\aboverulesep
          \global\let\@dashdrawstore\adl@draw
          \global\let\adl@draw\adl@drawiv}
  \cdashline{#1}
  \noalign{\global\let\adl@draw\@dashdrawstore
          \vskip\belowrulesep}}
\makeatother

\newcommand{\conceptulization}[1]{
    \textbf{\textcolor{orange}{#1}}
}
\newcommand{\stateTrust}[1]{
    \textbf{\textcolor{state_color}{#1}}
}
\newcommand{\trustee}[1]{
    \textbf{\textcolor{trustee_color}{#1}}
}
\newcommand{\trusteeAttributes}[1]{
    \textbf{\textcolor{trustee_attributes_color}{#1}}
}

\newenvironment{summary}{
    \vspace{0.3cm}
    \hrule 
    \vspace{0.15cm}
    \noindent{\textbf{Summary:} }
} {
    \vspace{0.15cm}
    \hrule
    \vspace{0.15cm}
}

\makeatletter
\if@todonotes@disabled

\else

\fi
\makeatother

\copyrightyear{2023}
\acmYear{2023}
\setcopyright{rightsretained}
\acmConference[CHI '23]{Proceedings of the 2023 CHI Conference on Human Factors in Computing Systems}{April 23--28, 2023}{Hamburg, Germany}
\acmBooktitle{Proceedings of the 2023 CHI Conference on Human Factors in Computing Systems (CHI '23), April 23--28, 2023, Hamburg, Germany}\acmDOI{10.1145/3544548.3581019}
\acmISBN{978-1-4503-9421-5/23/04}

\begin{document}

\title{What Do We Mean When We Talk about Trust in Social Media? A Systematic Review}  

\author{Yixuan Zhang}  
\affiliation{
    \institution{Georgia Institute of Technology}
    \city{Atlanta}
    \state{GA}
    \country{USA}}
\email{yixuan@gatech.edu}
\orcid{0000-0002-7412-4669}

\author{Joseph D Gaggiano}  
\affiliation{
    \institution{Georgia Institute of Technology}
    \city{Atlanta}
    \state{GA}
    \country{USA}}
\email{jgaggiano@gatech.edu}
\orcid{0000-0002-9740-4989}

\author{Nutchanon Yongsatianchot}  
\affiliation{
    \institution{Northeastern University}
    \city{Boston}
    \state{MA}
    \country{USA}}
\email{yongsatianchot.n@northeastern.edu}
\orcid{0000-0003-1332-0727}

\author{Nurul M Suhaimi}  
\affiliation{
    \institution{Universiti Malaysia Pahang}
    \state{Pahang}
    \country{Malaysia}}
\email{nmsuhaimi@ump.edu.my}
\orcid{0000-0002-4318-5805}

\author{Miso Kim}  
\affiliation{
    \institution{Northeastern University}
    \city{Boston}
    \state{MA}
    \country{USA}}
\email{m.kim@northeastern.edu}
 
\author{Yifan Sun}  
\affiliation{
    \institution{William \& Mary}
    \city{Williamsburg}
    \state{VA}
    \country{USA}}
\email{ysun25@wm.edu}
\orcid{0000-0003-3532-6521}

\author{Jacqueline Griffin}  
\affiliation{
    \institution{Northeastern University}
    \city{Boston}
    \state{MA}
    \country{USA}}
\email{ja.griffin@northeastern.edu}
\orcid{0000-0001-7729-1748}

\author{Andrea G Parker}  
\affiliation{
    \institution{Georgia Institute of Technology}
    \city{Atlanta}
    \state{GA}
    \country{USA}}
\email{andrea@cc.gatech.edu}
\orcid{0000-0002-2362-7717}
 
\renewcommand{\shortauthors}{Zhang et al.}

\begin{abstract} %150 word max; now 150 words
Do people trust social media? If so, why, in what contexts, and how does that trust impact their lives? Researchers, companies, and journalists alike have increasingly investigated these questions, which are fundamental to understanding social media interactions and their implications for society. However, trust in social media is a complex concept, and there is conflicting evidence about the antecedents and implications of trusting social media content, users, and platforms. More problematic is that we lack basic agreement as to what trust means in the context of social media. Addressing these challenges, we conducted a systematic review to identify themes and challenges in this field. Through our analysis of 70 papers, we contribute a synthesis of how trust in social media is defined, conceptualized, and measured, a summary of trust antecedents in social media, an understanding of how trust in social media impacts behaviors and attitudes, and directions for future work.
\end{abstract} 

\begin{CCSXML}
<ccs2012>
   <concept>
       <concept_id>10003120.10003121</concept_id>
       <concept_desc>Human-centered computing~Human computer interaction (HCI)</concept_desc>
       <concept_significance>500</concept_significance>
       </concept>
   <concept>
       <concept_id>10003120.10003130</concept_id>
       <concept_desc>Human-centered computing~Collaborative and social computing</concept_desc>
       <concept_significance>500</concept_significance>
       </concept>
 </ccs2012>
\end{CCSXML}

\ccsdesc[500]{Human-centered computing~Human computer interaction (HCI)}
\ccsdesc[500]{Human-centered computing~Collaborative and social computing}
\keywords{systematic review, social media, trust}


\maketitle

\section{Introduction \& Background}
\label{sec:intro}  
Social media~\footnote{In this paper, we define social media as any Internet-based platform that allows the creation and exchange of content, usually using either mobile or web-based technologies~\cite{margetts2015political}, such as Facebook and Twitter, etc.} has become an essential part of people's daily routines. As of 2022, around 4.48 billion people use social media worldwide; more than double the 2.07 billion people who used these platforms in 2015~\cite{dean2022social}. A 2020 study showed that the average adult spent 180 minutes a day on social media, compared to 90 minutes in 2012~\cite{hiley2021how}. This large increase in social media use is particularly consequential because it influences people's attitudes, decision-making, and behaviors~\cite{flintham2018falling, baxter2019scottish, zhang2022shifting}. A growing body of research has characterized how engagement on social media platforms, such as Facebook and Twitter, impacts health behaviors~\cite{hether2014s, woko2020investigation} and political engagement~\cite{carlisle2013social}, and how it influences attitudes towards societal issues, such as climate change~\cite{xenos2014great} and vaccine safety~\cite{jennings2021lack}. Given the centrality of social media in people's daily life and the fact that our attitudes and behaviors are shaped by it, people have to decide which social media platforms, information, and users they can and should trust, with the understanding that not all \textit{trustees} (i.e., who or what is being trusted) are credible online. 

As such, trust in social media, and a lack thereof, have become key topics of focus for researchers, businesses, and journalists alike. Over the past two decades, we have seen an increasing number of news articles~\cite{CBS2018trust, sutton2018eroding}, reports (e.g., Twitter Investigation Report~\cite{DFS2020Twitter}), and scholarly publications investigating how, why, and with what implications people do or do not trust social media~\cite{hsieh2021can, ray2021young, salmon2021axios, suhaimi2022investigating, zhang2020understanding}. As this body of work has grown, several challenges have simultaneously arisen. First, defining the concept of trust, characterizing how it is formed, and demonstrating how it impacts people is a complex endeavor. This complexity is evidenced by the lack of consensus on the meaning of ``trust'' in the context of social media'', the conflicting results from prior research regarding who and what strengthens trust in social media, and the diverse impacts of trust on people's behaviors and attitudes~\cite{hether2014s, cheng2020encountering, zhang2022shifting}. % Despite a large body of work in this area, we lack a synthesized understanding of the learnings, challenges, open questions, and implications for future work across this body of research. Such a synthesis is essential to support rigorous and impactful research moving forward.  
Beyond trust in social media, trust is an important concept that has been studied in HCI and other fields focused on information and technology. Prior reviews have systematically studied trust in several contexts, such as health information websites~\cite{kim2016trust}, mobile commerce~\cite{sarkar2020meta}, and artificial intelligence (AI)-enabled systems, including chatbots, robots, automated vehicles, and nonembodied algorithms\cite{bach2022systematic, kaplan2021trust}. Yet, we see that the antecedents of trust uncovered in these reviews vary across contexts. For example, a systematic review focused on mobile commerce~\cite{sarkar2020meta} identified a set of factors that influence users' trust, including service quality and perceived ease of use of the platform. On the other hand, in a review focused on trust in AI~\cite{kaplan2021trust}, the authors summarized a set of AI-related antecedents of trust, which included the level of automation used in the AI system and model performance. These existing reviews support the fact that trust is highly contextualized. As such, trust in social media is worthy of discussion and research in its own right. Despite a large body of work focused on trust in social media, we lack a synthesized understanding of the learnings, challenges, open questions, and implications for future work across this body of research. Such a synthesis is essential to support rigorous and impactful research moving forward. 
% \todo{Added more background of other reviews on trust to address R3's comments that ask for more related work on trust.} 

To address these challenges, we conducted a systematic review of prior research focused on trust in social media. Our work aimed to address four open research questions. 

First, trust is multifaceted and is a notoriously complex and nuanced concept~\cite{lewis1985trust}, and the understanding of trust also largely depends on the domain of study~\cite{harrison2001trust, tang2015trust}~\footnote{Scholars such as Blomqvist~\cite{blomqvist1997many} and Gefen et al.~\cite{gefen2003trust} have complied a list of diverse meanings of trust across the fields of social psychology, philosophy, economics, law, and marketing.}. %For example, trust in economics often seeks to explain the differences between people's actual behavior and their desire to maximize their own utility. In this context, trust is often framed using a rational choice perspective~\cite{morgan1994commitment} that presumes if trustees are perfectly honest and doing their best to fulfill their commitments, trust is facilitated~\cite{blomqvist1997many}. On the other hand, philosophers emphasize attitudes and beliefs when defining trust (often unconscious; as being part of everyday life)~\cite{blomqvist1997many}. For example, Plato and Aristotle, described trust in the context of cooperation between people, often clustered with other interpersonal components, such as friendship and diplomacy~\cite{baier1986trust, hertzberg1988attitude}. %Given that trust is highly contextualized, we argue that trust in social media is worthy of discussion and research in its own right. 
Moreover, the terms ``distrust'' and ``mistrust''---two highly-related concepts to trust---add another layer of complexity to the understanding of trust. The argument is whether or not trust and distrust should be considered as one concept with two extremes or as two distinct concepts~\cite{cheng2020encountering, harrison2001trust}. Therefore, there is a need for conceptual clarity in the context of trust in social media, to help focus, guide, and support comparisons between future work. Accordingly, our first research question is: \textbf{RQ1. How has existing work conceptualized and defined \textit{trust concepts} (that include the concepts of trust, distrust, and mistrust) in social media?}

Second, a topic closely related to the definitions of trust is the measurements of it. ``Good'' measurements should capture and align with the definition of what is being measured in order to produce rigorous research~\cite{harrison2001trust, kaplan2017conduct}. Accordingly, our next research question is: \textbf{RQ2. How are \textit{trust concepts} in social media measured empirically?}

Third, prior research in HCI and other fields has sought to determine the \textit{antecedents of trust}---that is, factors that can influence trust~\cite{harrison2001trust, sollner2013we}. However, there is a lack of a synthesized understanding regarding these antecedents of trust in social media. Specifically, it is unclear the extent to which the antecedents of trust identified in other contexts can be applied to social media, and what other unique antecedents may exist in this context. This lack of understanding constrains our ability to examine how to address the issues of trust surrounding social media (e.g., how to create interventions for social media that combat misinformation). Therefore, our third research question is: \textbf{RQ3. What are the antecedents of \textit{trust concepts} in social media?}
 
Lastly, it is also crucial to understand how trust in social media impacts people---the consequences of trust in social media. Existing work has explored several ways in which trust in social media influences people's behaviors and attitudes~\cite{flintham2018falling, baxter2019scottish}. For example, prior work has found that people who held a higher level of trust in health information on social media tended to engage in health behaviors (e.g., preventive behaviors, like avoiding crowds during the COVID-19 pandemic)~\cite{wu2022exploring}. On the other hand, other work has shown that trust in the information found on social media negatively influences people's health behaviors (e.g., by increasing vaccine hesitancy)~\cite{clark2022role}. Given the varied and conflicting evidence regarding the consequences of trust,  there is a need for a comprehensive synthesis and analysis of this body of literature regarding how trust impacts people's behaviors and attitudes. Accordingly, our last research question is: \textbf{RQ4. What are the consequences of \textit{trust concepts} in social media?}

In summary, given the complexity and ambiguity of the concept of trust in the context of social media, there is a need for a systematic review that maps the landscape of literature in this field. In conducting such a review, our work seeks to identify common themes and challenges in terms of the definition, conceptualization, measurement, antecedents, and consequences of trust in social media. To ground our discussion for the remainder of the paper, here we list a few key terms that will be used throughout this paper. \textit{Trustee} refers to the party or the object to be trusted, whereas \textit{trustor} refers to the party that makes the decision whether or not to trust~\cite{sekhon2014trustworthiness}. \textit{Trust concepts} include the concept of trust, distrust, and mistrust.

\textbf{Contributions.} Our work seeks to \ul{demystify}, \ul{disambiguate}, \ul{operationalize}, and \ul{defamiliarize} the commonly-used terms of trust, mistrust, and distrust in social media. In so doing, we contribute to the development of descriptive power in this research area (i.e., our ability to study, make sense of, and report on issues related to trust in social media) and rhetorical power~\cite{halverson2002activity} (i.e., communicate as a research community about the key qualities of \textit{trust concepts}). In so doing, we provide ``building blocks'' for future work that aims to make theoretical contributions (e.g., developing frameworks, models, and theories around trust in social media) and/or empirical contributions through investigations of trust surrounding social media. Specifically, we contribute: \textbf{(1)} an understanding of current methodological trends in prior research on \textit{trust concepts} in social media, such as research methods used, characteristics of studied populations, and social media platforms examined. Characterizing these trends will help guide the design of future studies, including work that investigates the understudied dimensions of this research space; \textbf{(2)} a comprehensive investigation regarding the definitions, conceptualizations, and measurements of \textit{trust concepts} in social media. Clarifying these fundamental concepts will help produce rigorous research. By laying out the definitions and conceptualizations, we provide conceptual contributions that can help guide research to better describe the meaning of trust within social media research and name key aspects of the conceptual structure; \textbf{(3)} a synthesized understanding of the antecedents of trust in social media, as well as the consequences of trust in social media on people's behaviors and attitudes. This understanding provides important insights into strategies and interventions for increasing the trustworthiness of social media information, users, and platforms. In short, we hope that this review will serve as an important resource for guiding future studies focused on trust in social media.

% Furthermore, our findings also bring attention to the crucial need for ethics-focused research that aims to prevent the manipulation of the public's trust as it relates to social media.

%%%%%%%%%%%%%%%%%%%%%%%%%%%% Method Section %%%%%%%%%%%%%%%%%%%%%%%%%%%%%%%%%%%%%%%%%%%%%%%%%%%%%%
\section{Method}    
\begin{figure*}[h]
  \centering
  \includegraphics[width=0.9\linewidth]{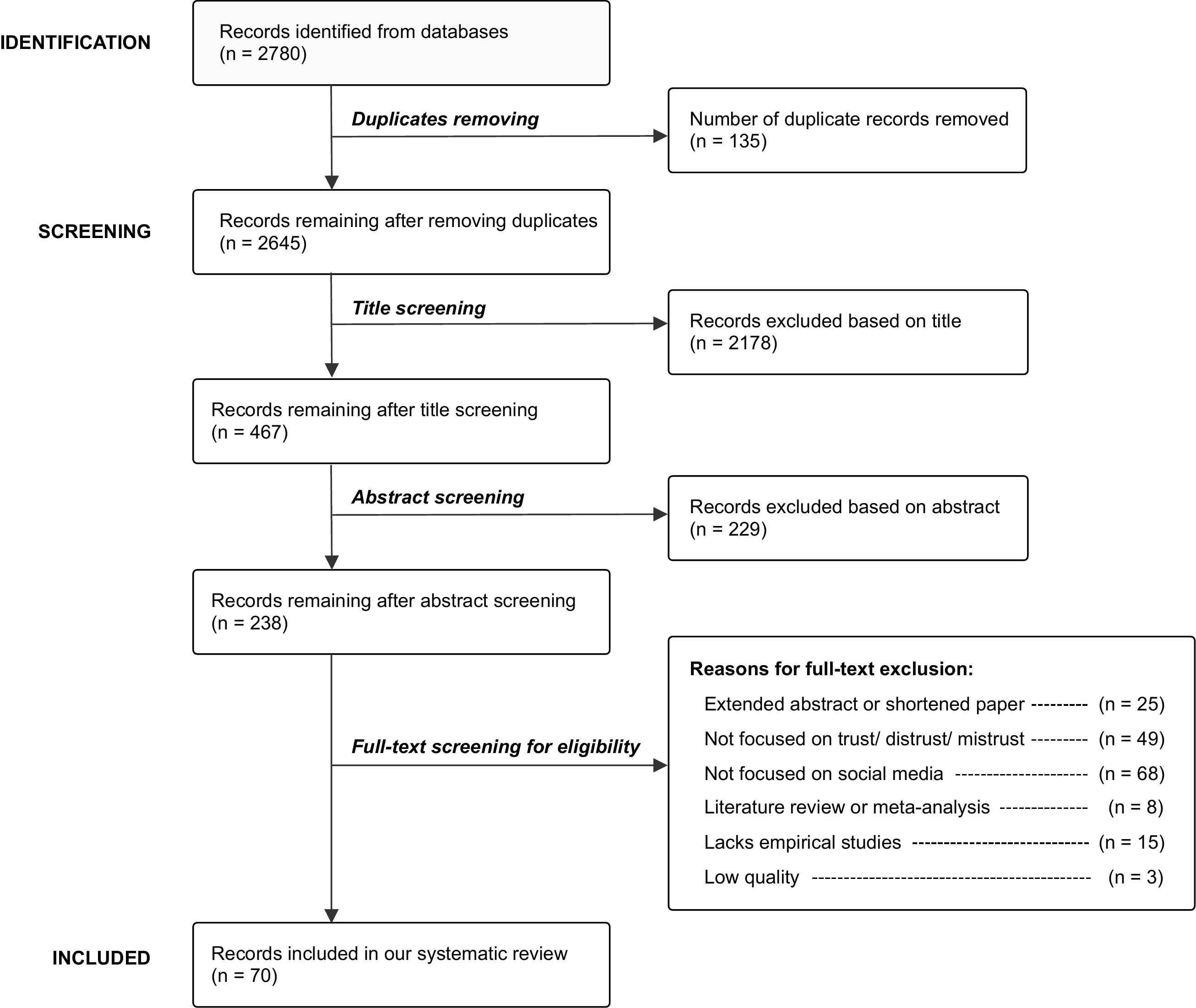}
  \caption{A PRISMA flow diagram~\cite{page2021prisma} that describes the processes and steps of our review, as well as the number of records returned.} 
  \label{fig:prisma_diagram}
\end{figure*}

Our review includes the following steps: 1) developing the search strategy and performing the literature search, 2) conducting a title screening, 3) performing an abstract screening, 4) completing a full-text screening, 5) running a quality assessment, and 6) extracting and analyzing the data.  \autoref{fig:prisma_diagram} provides an overview of our searching process and related search results, adapted from a widely used flow diagram template, called the Preferred Reporting Items for Systematic Reviews and Meta-Analyses (PRISMA)~\cite{page2021prisma}.

\subsection{Search Strategy}
\label{subsec:mothod_search_strategy}

\textbf{Databases:} Our database search was conducted from October to November 2021. A total of six databases were included: ACM Digital Library, IEEE Xplore, SAGE Journals, ScienceDirect (Elsevier), Taylor \& Francis Online (Tandfonline), and Emerald Insight. These databases were chosen because they are well-known digital libraries within the fields of social sciences and social media and contain a large amount of research related to trust in social media. 

\noindent
\textbf{Search Scope and Strings:} We used the Boolean ``OR'' to combine alternate terms (case insensitive) within each scope regarding the social media platform(s) and used ``AND'' to join the major trust concepts. %These social media platforms were selected because of their prevalence in past research and their dominance in industry~\cite{pew2021socialmedia}. 
Specifically, we apply the following search string to the metadata (including the title, keywords, and abstract) of the papers in each database:
\begin{displayquote}
 ("social media" \textbf{OR} "facebook" \textbf{OR} "twitter" \textbf{OR} "instagram" \textbf{OR} "reddit" \textbf{OR} "youtube" \textbf{OR} "whatsapp" \textbf{OR} "wechat") \textbf{AND} ("trust" \textbf{OR} "distrust" \textbf{OR} "mistrust")
\end{displayquote}

\subsection{Title Screening}
\label{subsec:method_title_screening}
We screened all search outcomes using a two-step process: a title-abstract review and a full-text review. In the first screening step, one researcher screened the titles and abstracts of each of the papers, dissociating relevant and irrelevant papers based on their congruity to our research questions outlined. Although there were 2780 results identified in our search, some articles were populated more than once across multiple search strings, resulting in 135 duplicate articles being removed from our list. Thus, a total of 2,645 papers were reviewed. Titles were considered relevant if they included keywords relating to \textit{trust concepts} and social media. Overall, 2178 papers were excluded based on their title, meaning a total of 467 papers held relevant titles to our research questions and were to be further screened.   

\subsection{Abstract Screening}
\label{subsec:method_abstract_screening}
One researcher read and analyzed the abstracts of the 467 articles to discern their relevance to the topic space. In other words, if the abstract established that the paper was related to social media and \textit{trust concepts}, the paper was included in the full-text analysis. There were 229 articles removed during the abstract review, resulting in 238 relevant articles remaining for the full-text screening.

\subsection{Full-text Screening}
\label{subsec:method_fulltext_screening}
Two researchers then closely examined the remaining 238 articles to determine the final inclusion of papers according to inclusion and exclusion criteria constructed by the researchers.

\textbf{Inclusion criteria.} 
Studies that satisfied all of the following inclusion criteria were included: 1) the paper is based on empirical research, and 2) the paper examines \textit{trust concepts} in social media.  

\textbf{Exclusion criteria.} 
Studies that met at least one of the following exclusion criteria were removed: 1) the article was not written in English, 2) the article was not peer-reviewed, 3) the article was an abstract, or an extended abstract, 4) the article was secondary research, such as review papers.

The research team held regular meetings (e.g., weekly meetings and day-to-day online communication) to validate the inclusion and exclusion of articles.

\subsection{Quality Assessment} 
\label{subsec:method_quality_assessment}
We then assessed the quality of each article of the remaining papers in our corpus via \textit{critical appraisal tools}, which provide analytical evaluations of the quality of each study through free and online checklists and worksheets~\cite{crombie2022pocket, katrak2004systematic}. Each article was evaluated by two researchers independently. Any disagreement regarding the quality of the article was discussed until consensus was achieved. Three papers were marked as being ``poor'' based on our critical appraisal tools and were removed from the corpus of our review. After all the inclusion and exclusion criteria processes, a total of 70 papers were included in our review~\footnote{For a complete list of included papers and a summary of these papers, please refer to the supplemental material.}.  

\subsection{Data Extraction and Analysis}
\label{subsec:method_extraction_analysis}
To streamline and systematize the analysis, the research team iteratively developed a protocol to guide data extraction. Our protocol collected both general information about the study design, as well as specific information relevant to our research questions, which aligns with review protocols suggested by prior work~\cite{levett2020systematic}. Specifically, we recorded the following data: the author(s), year of publication, study location in which the research was conducted, research questions (and hypotheses, if any), methodology and sample size, characteristics of the studied population, social media platform studied, the definition of \textit{trust concepts}, measurements used in each study, antecedents of trust, and consequences of trust. To assess the measurement of trust, we extracted all the questions related to \textit{trust concepts }asked in prior work that included both quantitative and qualitative studies. We determine the antecedents and consequences of trust based on the research questions and hypotheses (along with the rationale for proposing their hypotheses) provided in our surveyed papers.

Two researchers independently extracted the information from each paper and recorded it into spreadsheets.  A high inter-rater reliability (IRR), with a Cohen's kappa ($k$) coefficient of 0.81, was achieved. Then the two researchers compared and discussed discrepancies in data extraction to achieve a consensus.

We then took several explanatory approaches for data analysis, such as frequency analysis. Specifically, variables with similar dimensions across papers were clustered, and the total counts and percentages of articles in each cluster were calculated. Some dimensions required categorization beyond one level, so nested clusters were used to designate these cases. These clusters are visualized in tables, line charts, heat maps, and Sankey diagrams in our results (\autoref{sec:results}).

%%%%%%%%%%%%%%%%%%%%%%%%%%%% Results Section %%%%%%%%%%%%%%%%%%%%%%%%%%%%%%%%%%%%%%%%%%%%%%%%%%%%%%
\section{Results}
\label{sec:results}

\begin{figure}[h]
  \centering
  \includegraphics[width=\linewidth]{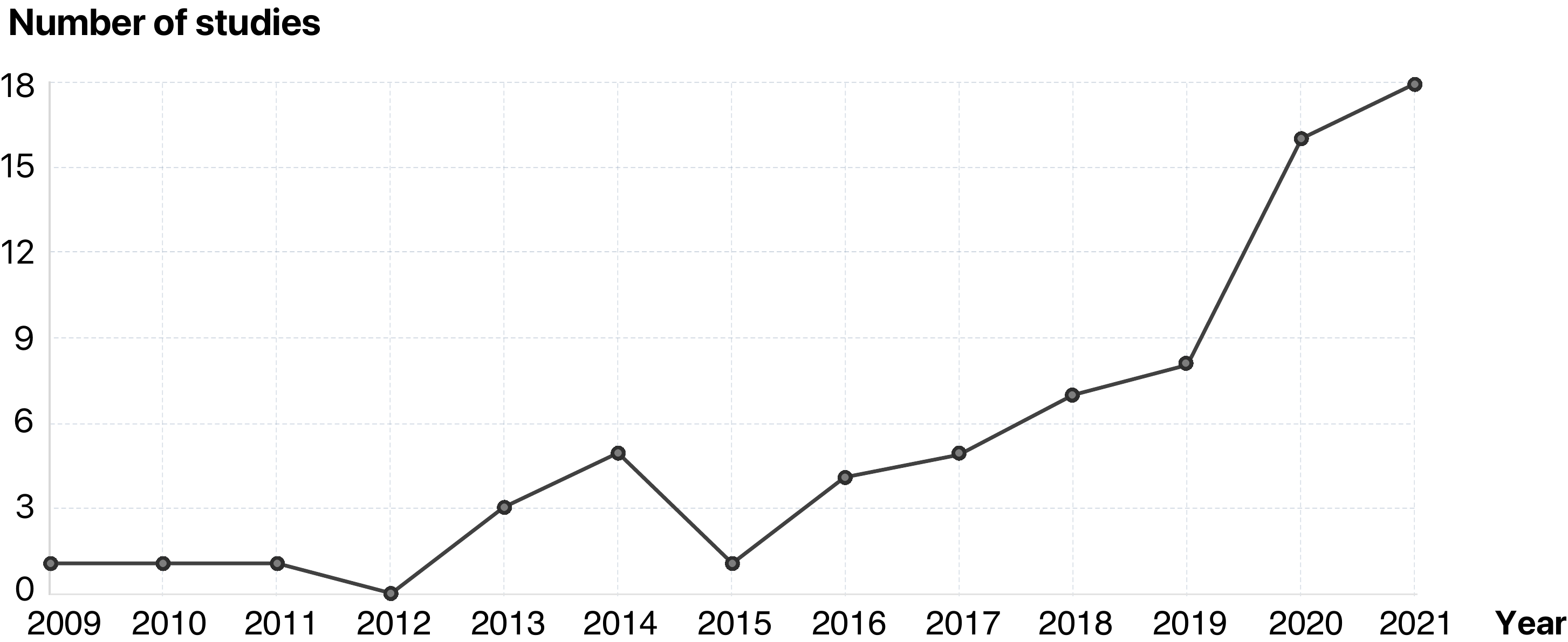}
  \caption{A time-series chart shows the number of studies that examined trust in social media in our corpus from 2009 to 2021.} 
  \label{fig:publication_year}
\end{figure}

Our results revealed that the research topic focused on trust in social media has become more popular over time, as shown in \autoref{fig:publication_year}. Most of the papers in our corpus were published in the last five years between 2017 and 2021 (n=54, 77\%); the rest were from 2009 to 2016 (n=16, 23\%).

Below we present findings related to the study design of our surveyed papers to overview the current status of research (\autoref{subsec:study_design}). Then, we describe the definitions and conceptualizations of trust in social media (\autoref{subsec:definition}), the measurements of trust (\autoref{subsec:measurements}), the antecedents of trust (\autoref{subsec:antecedents}), the consequences of trust (\autoref{subsec:impact}), and distrust and mistrust in social media (\autoref{subsec:distrust_mistrust}).  

\subsection{Study Design}  
\label{subsec:study_design} 

\subsubsection{Research Methods \& Sample Size} 

\begin{table*}[h!]
   \caption{An overview of research methods used among studies in our corpus and a summary of sample size, including the minimum and maximum number of participants ([Min, Max]), Median, Interquartile Range (IQR), Mean, and Standard Deviation (SD).} 
   \label{tab:research_methods}
   \begin{tabular}{lcrrrrrr}
        \toprule   
                &   & \multicolumn{6}{c}{\textbf{Sample Size}} \\ 
% \begin{imageonly}\cdashline{3-8}\end{imageonly}
                &   & \multicolumn{6}{c}{{----------------------------------------------------------------------------------------------}} \\
        \textbf{Methods}& \textbf{Number of studies*} & \multicolumn{1}{c}{[Min, Max]}  & Median & IQR & & Mean & SD \\
        \midrule 
            Survey            & 61  &[37, 6383]  & 402  & 493 & & 402 & 1034 \\   
            Experiment        & 13  &[39, 2026]  & 203  & 434 & & 203 & 592\\   
            Interview         & 6   &[4, 71]     & 9    & 9   & & 9   & 28\\  
        \midrule
        \multicolumn{4}{l}{
        \footnotesize\textit{*Some work used multiple methods.} } \\
        \bottomrule
    \end{tabular} 
\end{table*}

The vast majority of surveyed articles used a single research method (n=62, 89\%), while the remaining used multiple methods (n=8, 11\%). \autoref{tab:research_methods} provides the distribution of the specific research methods used within the articles in our corpus, with statistics of the sample size data provided. Overall, survey methods were the most popular approach, followed by experimental studies; very few qualitative studies were conducted.

\subsubsection{Time Periods of Study}  
The majority of studies (n=67, 96\%) were single time-point research (i.e., research carried out over a single period of time). Three studies (n=3, 4\%) gathered data in two waves (i.e., time points) of measurement.

\subsubsection{Social Media Platforms} 

\autoref{tab:platform} shows an overview of the social media platform(s) examined among our surveyed studies. Most studies (n= 43, 61\%) focused on a specific platform(s), while the rest addressed social media as a general concept (n=27, 39\%), utilizing questions and delineating findings that pertained to social media in general. Among the specific platform(s) examined, Facebook was by far the most commonly-reported individual platform, followed by Twitter, WeChat, Instagram, Weibo, YouTube, Snapchat, and others.  
%(n=33). Twitter (n=8), WeChat (n=4), Instagram (n=3), Weibo (n=2), Youtube (n=2), and Snapchat (n=2) were reported considerably less among the papers in our corpus. Other platforms examined include WhatsApp, MySpace, DingXiangYuan, Renren, etc. (each n=1). %and StudiVZ 

\begin{table}[H]
  \caption{An overview of the social media platforms examined.}
  \label{tab:platform}
  \begin{tabular}{lrrl} 
        \toprule
        \textbf{Platform} & \textbf{Number of}     & \textbf{Percentage}\\ 
        & \textbf{studies}     & \textbf{\,\,\,\,of studies*\% }\\ 
        \midrule 
            General    & 27 & 39\% & \mybar{.39}\\   
            Facebook   & 13 & 19\% & \mybar{.19}\\   
            Twitter    & 8  & 11\% & \mybar{.11}\\   
            WeChat     & 4  & 6\%  & \mybar{.06}\\   
            Instagram  & 3  & 4\%  & \mybar{.04}\\
            Weibo      & 2  & 3\%  & \mybar{.03}\\
            YouTube    & 2  & 3\%  & \mybar{.03}\\
            Snapchat   & 2  & 3\%  & \mybar{.03}\\ 
            Others **   & 7  & 10\%  & \mybar{.1}\\
        \midrule
        \multicolumn{4}{l}{\footnotesize\textit{*Total \% is more than 100\% as some papers used multiple methods.}} \\ 
        \multicolumn{4}{l}{\footnotesize\textit{** If a platform was only examined by one study, then this platform}} \\ 
\multicolumn{4}{l}{\footnotesize\textit{was labeled as ``others''.}} \\
        \bottomrule
    \end{tabular} 
\end{table}

\subsubsection{Study Population} 

\begin{table*}[h]
  \caption{Number of studies (and associated percentages \%) that reported specific demographic characteristics of participants.} 
  \label{tab:demographic}
  \begin{tabular}{llrrl}
    \toprule
    \textbf{Characteristics}    & \textbf{Options} & \textbf{Number} & \textbf{Percentage (\%) } \\ 
                                &  & \textbf{of studies} & \textbf{of studies*} \\
    \midrule
    Sex or Gender  & Female \& Male           & 56 & 80\% & \mybar{.80}\\
                   & Women \& Men \& Others   & 5  & 7\% & \mybar{.07}\\
                   & Not reported (neither sex nor gender) & 9 & 13\% & \mybar{.13}\\
    \midrule  
    Race       & African-American & 13 & 19\%  & \mybar{.19}\\
               & American Indian  & 5  &  7\%\ & \mybar{.07}\\
               & Asian            & 8  & 11\%  & \mybar{.11}\\
               & White            & 13 & 19\%  & \mybar{.19}\\
               & Other            & 1  & 1\%   & \mybar{.01}\\
               & Not reported     & 53 & 76\%  & \mybar{.76}  \vspace{0.1cm}\\
    Ethnicity  & Latino/Hispanic  & 9 & 13\% & \mybar{.13}\\ 
               & Not reported     & 61 & 87\%& \mybar{.87}\\
    \midrule
    Education   & High school or less & 18 & 26\% & \mybar{.26} \\
                & Associate degree & 3  & 4\%  & \mybar{.04}\\
                & Vocational degree & 4  & 6\%  & \mybar{.06}\\
                & Diploma degree    & 6  & 9\%  & \mybar{.09}\\
                & Some college      & 10 & 14\% & \mybar{.14}\\
                & Bachelor degree   & 22 & 31\% & \mybar{.31}\\
                & Graduate degree   & 19 & 27\% & \mybar{.27}\\ 
                & Other             & 1  & 1\%  & \mybar{.01}\\
                & Not reported      & 26 & 37\% & \mybar{.37}\\
    \midrule
    Geo-location & North America     & 23 & 33\% & \mybar{.33}\\   
                 & Asia              & 19 & 27\% & \mybar{.27}\\   
                 & Europe            & 12 & 17\% & \mybar{.17}\\   
                 & South America     & 2  & 3\%  & \mybar{.03}\\   
                 & Australia         & 1  & 1\%  & \mybar{.01}\\
                 & Multiple locations& 5  & 7\%  & \mybar{.07}\\
                 & Not specified     & 8  & 11\% & \mybar{.11}\\ 
    \midrule 
    Age          & Under 18 years old    
                 & 9  & 13\% & \mybar{.13}\\ 
                 & Between 18 and 60 years old    & 55 & 79\% & \mybar{.79}\\ 
                 & Above 60 years old       & 19 & 27\% & \mybar{.27}\\ 
                 & Not reported                             & 15 & 21\% & \mybar{.21}\\   
    \midrule
    Income level** & Reported     & 17 & 24\% & \mybar{.24}\\   
                   & Not reported & 53 & 76\% & \mybar{.76}\\ 
    \midrule
    Political ideology**  & Reported      & 6  & 9\%  & \mybar{.09}\\ 
                        & Not reported  & 64 & 91\% & \mybar{.91}\\ 
    \midrule
    \multicolumn{5}{l}{
        \footnotesize \textit{*Total percentage may be more than 100\% as many papers reported multiple characteristics options of each category.}
    } \\
    \multicolumn{5}{l}{ 
        \footnotesize \textit{**Low-level categorizations of income \& political affiliation were not included due to the inconsistency of reporting across countries and across papers.} 
    } \\ 
    \bottomrule
  \end{tabular}
\end{table*} 

\autoref{tab:demographic} presents an overview of the total number of studies (with percentages provided) that include each (sub)category of participants' characteristics. These demographics included sex and gender, race, ethnicity, and education. We report these characteristics to call attention to issues of diversity in HCI research~\cite{himmelsbach2019we}. We assert that it is critical to report demographic data, as individual characteristics may play a role in influencing trust as we will see later in \autoref{subsec:antecedents}. %We did not include low-level categorizations in other aspects, such as age, income, and political affiliation, due to the inconsistency of the categorization and reporting styles of these characteristics across papers.  

The most commonly-reported demographic data was geo-location, sex and/or gender, age, and education. In contrast, other dimensions, such as race and ethnicity, income, and political affiliation, were less commonly reported. In terms of the \textbf{study locations}, studies that examined populations in North America were the most popular, followed by Asia and Europe; few studies focused on people in South America and Australia. Of the papers that reported participants' \textbf{sex or gender}, sex (i.e., female, male) was far more assessed across studies than gender (i.e., women, men, etc.). Furthermore, for papers that reported the \textbf{race and/or ethnicity} of participants (n=17), most of them (15 out of 17) studied populations in the United States. In terms of \textbf{educational levels}, most studies included participants with bachelor's or graduate degrees compared to lower levels of education.

\begin{summary}
Overall, our findings suggest that research on trust in social media is an increasingly popular topic. Most studies in our corpus used a single research methodology, with surveys being most prominent, followed by experimental studies; few studies used qualitative research methods. Furthermore, almost all studies were single-time point studies, except for three survey studies that conducted two waves of research. In terms of the studied populations, the most commonly-reported demographic characteristics were age and sex or gender, whereas race, income level, and political affiliation were much less commonly reported. Additionally, more than half of the studies focused on specific platforms rather than social media as a general concept with Facebook being the most popular. 
\end{summary}

%%%%%%%%%%%%%%%%%%%%%%%%%%%% New Subsection %%%%%%%%%%%%%%%%%%%%%%%%%%%%%%%%%%%%%%%%%%%%%%%%%%%%%%
\subsection{Definitions and Conceptualizations of Trust}
\label{subsec:definition} 

Among the papers that focused on trust in social media, only about half of them (n=34, 49\%) provided clear definitions of trust concepts~\footnote{For cases where papers discussed trust definitions (oftentimes in related work) but did not explicitly specify the definition of trust that the research team adopted within their work (e.g., ``in this paper, we define trust as xxx'' or ``in our study context, trust refers to xxx''), we labeled them as having unclear definition(s) in their study context.}.

\subsubsection{General Trends and Flows of Trust Definitions}  
Overall, our findings suggest that the definitions of trust in the context of social media were ``adapted'' from a wide range of definitions offered by early trust scholars. One of the most commonly-adopted trust definitions (n=4) was from Mayer et al.~\cite{mayer1995integrative}, defining trust as \textit{``the willingness of a party to be vulnerable to the actions of another party based on the expectation that the other will perform a particular action important to the trustor, irrespective of the ability to monitor or control that other part.''} Other commonly-cited definitions of trust were from Moorman et al.~\cite{moorman1992relationships} who defined trust as \textit{``a willingness to rely on an exchange partner in whom one has confidence''} (n=2), and Gefen et al.~\cite{gefen2003trust} who defined trust as \textit{``a set of specific beliefs in the integrity, benevolence, ability, and predictability'' of a party} (n=2).

\begin{figure*}[h]
  \centering
  \includegraphics[width=\linewidth]{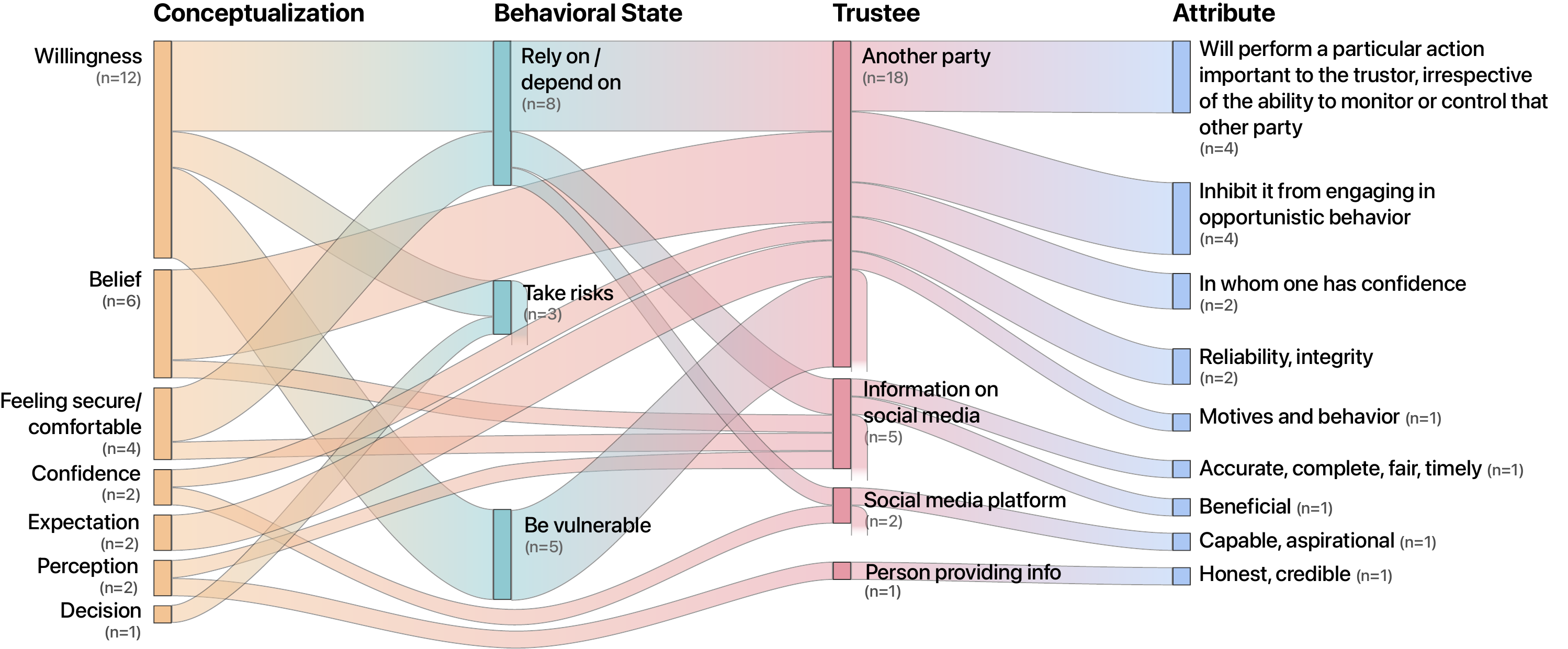}  
  \caption{A Sankey diagram that shows the flows of trust definitions of our surveyed papers. Note that the symbol \protect\includegraphics[height=2mm]{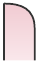} indicates the drop-offs (or flows without a target node).} 
  \label{fig:sankey_trust_definition}
\end{figure*}

\autoref{fig:sankey_trust_definition} is a Sankey diagram that illustrates the overall flows of trust definitions provided in our surveyed papers. This diagram helps (visually) understand some of the widely-used definitions of trust adopted by prior work and how trust in the context of social media is built upon the existing literature. 

To further understand how trust has been defined, we first consider how prior work has \textbf{conceptualized}~\footnote{A conceptualization can be considered an abstract, simplified view of trust.} the way in which trust manifests. Overall, trust has mostly been conceptualized as manifesting as a kind of \textit{willingness} on the part of the trustor (n=12), followed by a \textit{belief} (n=6), \textit{feeling} (n=4), \textit{confidence} (n=2), \textit{expectation} (n=2), \textit{perception} (n=2), and \textit{decision} (n=1). These conceptualizations were visually labeled as \includegraphics[height=2mm]{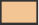} in \autoref{fig:sankey_trust_definition}.  
Following these conceptualizations of trust are terms describing \textbf{behavioral states} invoked by trust (visually encoded as \includegraphics[height=2mm]{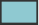} in \autoref{fig:sankey_trust_definition}). These states consist of phrases, such as \textit{``rely on/depend on''}, \textit{``take risks''}, or \textit{``be vulnerable''}. These conceptualizations and behavioral states create definitions, such as \textit{``willingness to depend on''}, \textit{``willingness to take risks''}, or \textit{``willingness to be vulnerable''}.  
Another component within each trust definition is the \textbf{trustee}, which emphasizes \emph{what} or \emph{whom} is being trusted (encoded as \includegraphics[height=2mm]{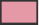} in \autoref{fig:sankey_trust_definition}). 
Most papers did not name a specific trustee, instead opting to use the phrase \textit{``another party''} as a general term to describe the trustee. 
Lastly, attributes that describe the \textbf{characteristics of the trustee} were also often provided within each definition (labeled as \includegraphics[height=2mm]{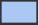} in \autoref{fig:sankey_trust_definition}). 
These attributes, in tandem with the trustee, create phrases, such as \textit{``information on social media is accurate, complete, fair, and timely''} and \textit{``another party that has reliability and integrity''}.

\subsubsection{Tailored Terms \& Definitions in The Context of Social Media} 

\begin{table*}[h]
% \small
  \caption{Tailored terms and definitions of \textit{trust concepts} in the context of social media (terms are sorted in alphabetical order): Similar to the color coding and dimensions described in \autoref{fig:sankey_trust_definition}, \conceptulization{text in orange} suggests the \conceptulization{conceptualization of trust} (e.g., being conceptualized as ``willingness'', ``perception'', ``belief'', and ``expectation''), \stateTrust{text in green} represents the \stateTrust{behavioral} states (wherever applicable) following the conceptualization of trust, \trustee{text in violet-red} indicates the \trustee{trustee}, and \trusteeAttributes{text in blue} shows the \trusteeAttributes{attributes of the trustee} (wherever applicable).}  
  \label{tab:tailored_trust_definition}
  \begin{tabular}{p{2.5cm}p{11.8cm}}
    \toprule   
    \textbf{Trust terms} & \textbf{Tailored definitions}\\
    \midrule 
    ``Information trust'' & \conceptulization{Belief} in the online \trustee{information} provided by an organization \trusteeAttributes{that is beneficial} to an individual~\cite{kim2021event}. \\  
    ``Media trust''       & The \conceptulization{perception} that the \trustee{information} is \trusteeAttributes{accurate, complete, fair, and timely}~\cite{kaye2020appsolutely}. \\  
    ``Mistrust in news'' &  Skeptical of the \trustee{news source} and defers the judgment~\cite{park2020global}.\\
    ``Online trust''      & The extent to which one \conceptulization{feels} secure and comfortable about \stateTrust{relying on} the \trustee{information on social media}~\cite{chauhan2020trustworthiness}. \\   
    ``Platform trust''    & A \conceptulization{psychological state} \trusteeAttributes{in which a person has favorable views of others' attributes}~\cite{geng2021effects}. \\  
    ``SNS trust''         & (Social media) users' \conceptulization{willingness} to \stateTrust{rely on} \trustee{others} \trusteeAttributes{in whom they have confidence}~\cite{phua2017gratifications}. \\  
    ``Source trust''      & The extent to which \trustee{the person providing the information} is \conceptulization{deemed} \trusteeAttributes{honest or credible}~\cite{peterson2021examining}.  \\ 
    ``Trustworthiness''  & \conceptulization{Perceived believability} of \trustee{information}~\cite{wang2013trust}.  \\ 
    ``Trust in SNSs''     & Consumer's \conceptulization{willingness} to trust the \trustee{SNSs}~\cite{kananukul2015building}  \\ 
    ``Trust'' & One's \conceptulization{willingness} to \stateTrust{depend on} \trustee{Facebook}~\cite{malik2016impact}. \newline
                An implicit \conceptulization{belief} that reflects online users' \conceptulization{confidence} in the \trustee{information platform}~\cite{cheng2020encountering}. \newline
                The extent to which you \conceptulization{feel secure and comfortable} \stateTrust{relying on} the \trustee{information}~\cite{chauhan2020trustworthiness}. \newline 
                The \conceptulization{belief} in \trustee{others} and in \trustee{their posted articles} on the \trustee{Facebook website}~\cite{yang2014people}.
                \\
    \bottomrule
  \end{tabular}
\end{table*}

In our corpus, 13 papers provided tailored terms and definitions of trust in the context of social media, as shown in \autoref{tab:tailored_trust_definition}. Overall, the conceptualizations among these tailored definitions were almost equally distributed between framing trust as a willingness, belief, feeling, and perception. The trustees included in each definition help specify whom or what is being trusted. 
% \pagebreak
\begin{summary}
Less than half of the papers in our corpus explicitly defined trust, and very few studies provided tailored definitions of trust in the context of social media, making the concept of trust ambiguous. Among the papers that did define it, trust has been most commonly conceptualized as willingness, with some conceptualizations framing trust as a belief, feeling, confidence, expectation, perception, or decision. 
\end{summary}

%%%%%%%%%%%%%%%%%%%%%%%%%%%% New Subsection %%%%%%%%%%%%%%%%%%%%%%%%%%%%%%%%%%%%%%%%%%%%%%%%%%%%%%
\subsection{Measurements of Trust}
\label{subsec:measurements}

Our results show that all but seven papers included detailed measurements of trust that they utilized in their studies, reflecting a high level of transparency in the research. Below, we present results on how well existing studies map their definitions to their measurements (to ensure construct validity), a low-level categorization of trustee(s) in the context of social media when measuring trust, and the dimensions of trust.  

\subsubsection{Construct Validity: Mapping Definitions to Measurements}  
Our results showed that overall, only 47\% of the studies mapped the definitions and/or conceptualizations of trust to the measurements. For example, as described in \autoref{subsec:definition}, ``willingness'' has been used the most often in our corpus to conceptualize trust. As such, measurements of trust in these specific papers intuitively should, at least in part, include a measurement of participants' willingness ~\cite{schoorman2007integrative}. However, surprisingly, none of the studies in our corpus specifically measured  ``willingness''. Instead of mapping trust measurements directly to the definitions specified, papers in our corpus tended to ask general questions when measuring trust, such as ``how much trust do you have on social media''. Similar issues are seen across the rest of the trust conceptualizations, as well. This misalignment of definitions to measurements diminishes the construct validity of trust (i.e., the extent to which the measure accurately assesses what it’s supposed to~\cite{bordens2014ebook}).

\subsubsection{Trustee: The Party or The Object to be Trusted}

The measurements of trust across our surveyed papers focused on the following trustees in the context of social media: 1) information from social media, 2) social media users/groups (like Facebook groups), 3) social media platforms, and in some cases, a combination of two or more.

Most measurements of trust specifically focused on a single trustee, except for eight studies that integrated multiple trustee measurements into one variable. Specifically, while these studies appear to use multiple survey questions to measure a \textit{single} trustee, these questions in fact ask about \textit{more than one} trustee. For example, in a set of questions intended to measure trust in a social media platform, some of these questions ask about the platform while others ask about users or information. Such scenarios introduce conceptual and empirical challenges. Specifically, these measures introduce threats to construct validity in that they are not precisely measuring the type of trust the research is setting out to assess. Indeed, individuals may have differing levels of trust in the platform versus users and information on social media. As such, a measure that sets out to evaluate platform trust but that asks about information and user trust as well is likely to yield a muddled and even inaccurate characterization of a respondent's level of trust towards the social media platform. We will further discuss the issue of integrating the measurement of multiple trustees into one variable in \autoref{sec:discussion}.

\subsubsection{Dimensions of Trust in Relation to the Trustee}
\label{subsubsec:dimensions_of_trust_by_trustee}

Trust is a multi-dimensional and complex construct~\cite{lewis1985trust}, indicating that trust might be better understood through multiple dimensions. However, our results show that most of the studies adopted single-item measurements (e.g., asking \textit{``how much do you trust information from social media''}), whereas far fewer used multi-item measurements. Broken down by trustee, 67\% of studies that focused on trust in information from social media, 88\% of studies that focused on trust in social media users, and 84\% of studies that focused on trust in social media platforms adopted single-item measurements when evaluating trust.  

\begin{figure*}[h]
  \centering
  \includegraphics[width=\linewidth]{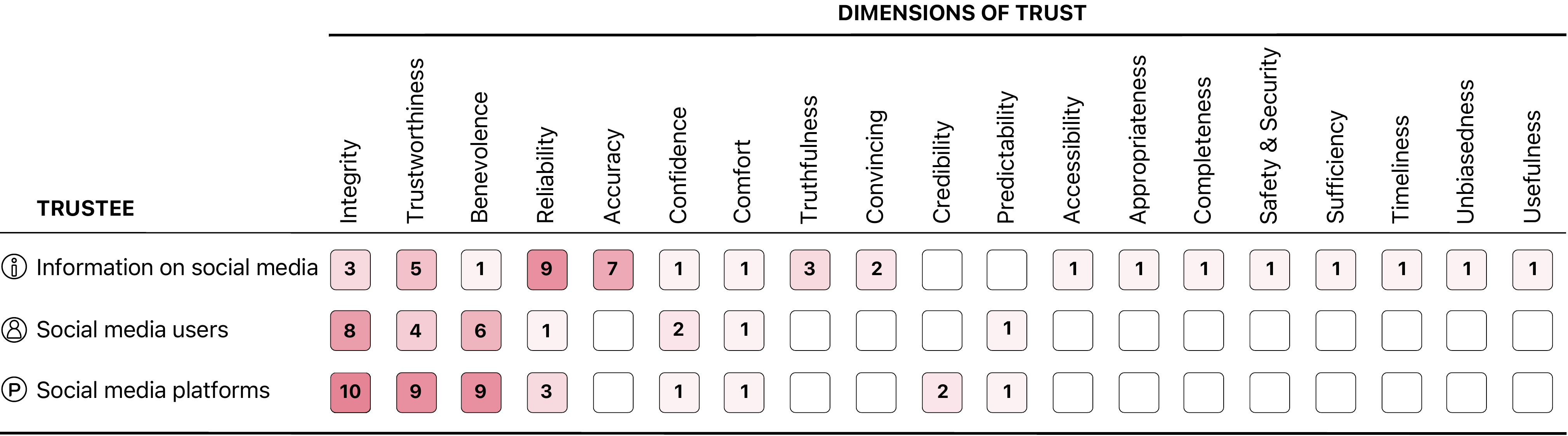}
  \caption{A heat map that visualizes the distribution of trust dimensions relative to the trustee. The darker the color, the greater number of studies that have examined the particular dimension of trust. The number inside the square \protect\includegraphics[height=3mm]{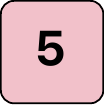} represents the total number of studies for each case.} 
  \label{fig:heatmap}
\end{figure*}

For studies that adopted multi-item measurements, we visualize the dimensions of trust relative to the three main trustees, as shown in \autoref{fig:heatmap}. We can see that there are a large variety of dimensions used to measure trust, ranging from integrity to usefulness. A few of these dimensions were directly borrowed from early trust literature, such as integrity (e.g., making good faith agreement) and benevolence (e.g., caring and being motivated to act in one's interest rather than acting opportunistically)~\cite{holmes1991trust}. 

On the other hand, in the context of social media, the dimensions used to measure trust have expanded, broadening in attempts to better characterize trust. Relatedly, the frequency of each trust dimension differs among trustees. For example, reliability and accuracy were often used when describing information, while integrity and trustworthiness were often used when describing social media platforms. Additionally, we also note that the measurements used in the studies overall do not align with their definitions.

\begin{summary}
We found a variety of issues within trust measurements in the context of social media:  1) most studies failed to map definitions to measurements, hurting construct validity; 2) many studies imprecisely specified the trustee (e.g., information on social media, social media users/groups, and/or social media platforms), further ambiguating trust in social media; and 3) most studies adopted a single-item measurement of trust compared to multiple-item, regardless of the fact that trust is a multi-dimensional phenomenon. 
\end{summary}

%%%%%%%%%%%%%%%%%%%%%%%%%%%% New Subsection %%%%%%%%%%%%%%%%%%%%%%%%%%%%%%%%%%%%%%%%%%%%%%%%%%%%%%
\subsection{Antecedents of Trust} 
\label{subsec:antecedents} 
\begin{figure*} %[htbp]  
  \centering
  \includegraphics[width=0.82\linewidth]{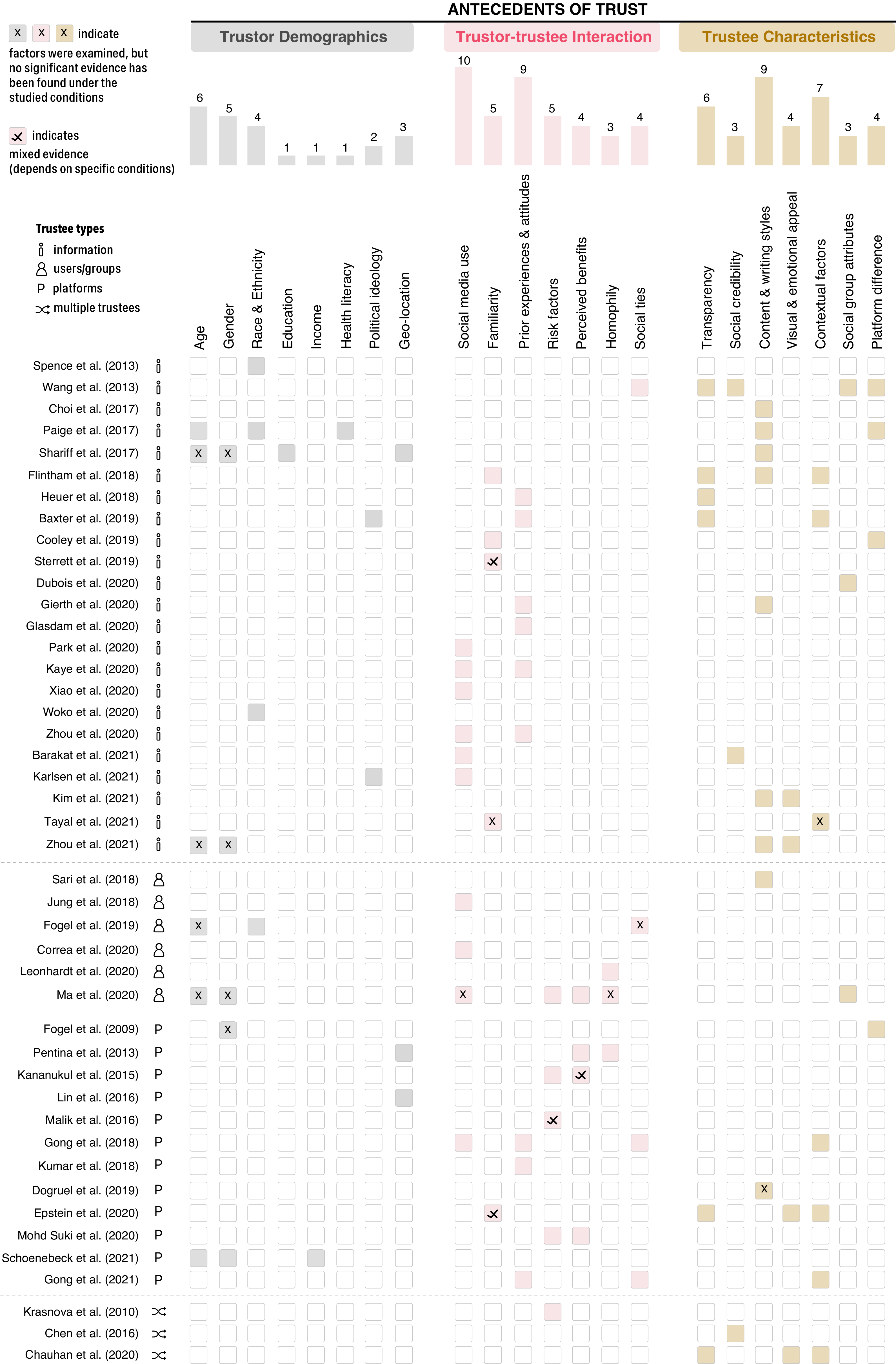}
  \caption{An overview of antecedents of trust in social media.} 
  \label{fig:antecedents}
\end{figure*}

As shown in \autoref{fig:antecedents}, antecedents of trust identified within the papers of our review were organized into 1) trust antecedents related to specific characteristics of the trustor, which we call \textit{demographic antecedents} (n=13); 2) trust antecedents related to the interaction and/or relationship between trustor and trustee, which we call \textit{trustor-trustee interaction antecedents} (n=29), and 3) trust antecedents related to specific characteristics of the trustee, which we term \textit{trustee-related antecedents} (n=21). We identified these antecedents of trust based on the hypotheses and assumptions described in the studies in our corpus. Note that the majority of studies examining the antecedents of trust in our corpus utilize observational or survey methodology, limiting any causal conclusions that can be drawn from the research. 
 
\subsubsection{Demographic Antecedents}
\label{subsubsec:demographic_antecedents} 
Below we describe \emph{demographic antecedents} examined by prior work, including age, gender, race and ethnicity, education, income, health literacy, political ideology, and geo-location. 

\textbf{Age.} Overall, there is mixed evidence on whether age impacts one's trust in information on social media, as well as trust in other social media users. Only two studies found that age is associated with people's trust. Specifically, one study found that older adults with low eHealth literacy (i.e., the skills to seek, find, appraise, and evaluate health information from electronic sources to address health concerns and make health decisions) had a higher level of trust in health information from Facebook~\cite{paige2017influence}. Likewise, another study found that older participants are more likely to trust Facebook~\cite{schoenebeck2021youth}. However, other studies did not find age to be a significant antecedent of trust in information from social media~\cite{shariff2017credibility, zhou2021building} nor to be a significant antecedent of trust in online groups on social media~\cite{fogel2019trust, ma2019when}. 

\textbf{Gender.} In general, our results show that gender was not a significant antecedent of trust in information~\cite{shariff2017credibility, zhou2021building}, trust in online social media groups~\cite{ma2019when}, nor trust in social media platforms~\cite{fogel2009internet}. One study did, however, find that transgender groups were more likely to distrust all social media platforms examined in their study~\cite{schoenebeck2021youth}. 

\textbf{Race and ethnicity.} Overall, studies in our corpus suggest that race and ethnicity were important antecedents of trust. Three studies focused on Black/African American populations, and one examined Asian populations. For example, Woko et al. found that Black/African Americans trusted social media for COVID-19 information significantly more than other populations~\cite{woko2020investigation}. Likewise, another study showed that Black/African Americans with low eHealth literacy had high perceived trust in YouTube and Twitter~\cite{paige2017influence}. Furthermore, Spence et al. found differences in perceived credibility depending on the race of the profile owner and the race of participants~\cite{spence2013gates}. Specifically, social media profiles with African American avatars were viewed as more trustworthy when posting information online, and African American participants rated all avatars as more trustworthy than White participants. On the other hand, one study suggested that Asian/Asian American participants had higher trust in social media prescription medication advertisements than participants of other races~\cite{fogel2019trust}. 

\textbf{Education.} One study examined the education levels of participants as a potential trust antecedent~\cite{shariff2017credibility} and found that as education increased, the perceived credibility of Tweets decreased.  

\textbf{Income.} One study examined income levels as a potential trust antecedent~\cite{schoenebeck2021youth}. This study found that lower-income groups (in the United States) trusted Facebook and Instagram more than other platforms to achieve fair resolutions after online harassment~\cite{schoenebeck2021youth}.

\textbf{Health literacy.} One study examined health literacy as a potential trust antecedent~\cite{paige2017influence}. This study found that older adults with low health literacy had high perceived trust in Facebook but low perceived trust in online groups~\cite{paige2017influence}.

\textbf{Political ideology.} Results from two studies suggest that personal political ideology or political affiliation influences participants' trust in information from social media~\cite{baxter2019scottish, karlsen2021social}. For example, Karlsen et al. found that Conservative supporters (a center-right party in Norway) found the news on social media less credible than Labour supporters (a center-left political party in Norway)~\cite{karlsen2021social}.  

\textbf{Geo-location.} Three studies found geo-location to be an antecedent of trust in social media platforms~\cite{lin2016health, pentina2013antecedents, shariff2017credibility}, indicating that cultural differences play a role in influencing trust in social media. For example, Lin et al.~\cite{lin2016health} found that youth in Hong Kong held a higher level of trust towards health-related information on social media, as compared to youth in the U.S. Another study~\cite{pentina2013antecedents} comparing American to Ukrainian participants found that Ukrainians who trusted a social media platform (i.e., Twitter) often transferred and generalized this trust to the profiles of commercial brands on that platform. Americans, however, did not.

\subsubsection{Trustor-trustee Interaction Antecedents}
\label{subsubsec:antecedents_trust_info}  

In addition to demographic antecedents, trust antecedents related to the interaction between the trustor and trustee were found across papers in our corpus---a category we termed \emph{trustor-trustee interaction antecedents}. For example, an individual's social media use may rely on that individual's (trustor's) relationship with the social media platform (trustee). These interactions and/or relationships involve both the trustor and trustee, differentiating this category of antecedents from the others. Within the papers in our corpus, trustor-trustee interaction antecedents include social media use, familiarity (with the trustee), prior experiences and attitudes, risk factors involved in interacting with the trustee, perceived benefits of interacting with the trustee,  homophily, and social ties. %Note that some of the trust antecedents may not be mutually exclusive. 

\textbf{Social media use.} Social media use, which is often measured by the amount of time someone spends on social media, is the most commonly-examined antecedent. Overall, there is mixed evidence on whether social media use influences trust in information on social media. Several studies indicate that the more frequently people used social media, the higher level of trust they had~\cite{barakat2021empirical, jung2018determinants, kaye2020appsolutely, xiao2021dangers, zhou2020internet}. However, two studies provided contradictory results, suggesting that higher usage of social media decreases trust~\cite{karlsen2021social, park2020global}. Additionally, another study found that individual levels of social media engagement did not significantly predict trust in social media groups~\cite{ma2019when}.    

In a similar vein, one study examined the concept of ``\textit{perceived critical mass}'', defined as the degree to which a person believes that most of their peers are also using the same platform~\cite{lou2000perceived}), as a trust antecedent. This study found that perceived critical mass was positively related to users' trust in the social media platform, WeChat~\cite{gong2018experienced}.  

\textbf{Familiarity.} Familiarity, in the context of our corpus, refers to the knowledge and recognition of the trustee. Similar to social media use, there is mixed evidence that familiarity with the trustee (e.g., an information source) influences people's trust. Two studies suggest that knowledge of and familiarity with a source, including social media contacts known personally, were essential trust antecedents~\cite{cooley2019effect, flintham2018falling}. However, indicators of trust are not necessarily antecedents, as familiarity with a platform is necessary but not sufficient for trust~\cite{epstein2020will}. This is because reputable but unfamiliar sources are likely to receive unnecessarily low trust scores from participants, as people were unacquainted and thus suspicious of these sources~\cite{epstein2020will}. Additionally, one study found that participants did not necessarily find information about COVID-19 to be more trustworthy if the post came from a familiar source, such as a friend or family member~\cite{tayal2021reliability}. 

\textbf{Prior experiences \& attitudes.} 
People's prior experiences with and attitudes towards the trustee influence their perceptions of that trustee and have been shown to be essential antecedents of trust in information found on social media~\cite{baxter2019scottish, gierth2020attacking, glasdam2020information, heuer2018trust, zhou2021building}. For example, Baxter et al. found that participants drew on personal experiences when determining the reliability of information found on social media~\cite{baxter2019scottish}. Similarly, Glasdam et al.'s qualitative study found that individuals' personal experiences influenced how people differentiate between COVID-19 misinformation and factual information~\cite{glasdam2020information}. 
%~\footnote{To some degree, individuals' prior experiences and attitudes are related to familiarity. How we group these factors (either familiarity or prior experiences and attitudes) largely depends on how existing work has termed their concepts.}

Prior experiences and attitudes have also been shown to influence people's trust in the social media platforms themselves. For example, Kumar et al. found that prior positive experiences on Facebook were positively related to trust in the platform~\cite{kumar2018interplay}. Likewise, two studies found that \textit{user satisfaction} with a social media platform was associated with increased trust in it~\cite{gong2018experienced, gong2021gender}. Furthermore, prior experiences are also highly related to an individual's \textit{perceptual bias} (i.e., assumptions made when assessing the degree of bias that information has on social media). One study found that participants perceived political news from social media as more biased than political news from other platforms, such as radio or TV, contributing to lower levels of trust in social media~\cite{kaye2020appsolutely}.

\textbf{Perceived benefits.} Higher levels of perceived benefits, such as perceived entertainment benefits and/or social support from social media, help increase trust in social media platforms. For example, one study found that individuals who believed that they received practical and social benefits from a social media platform had increased trust towards that platform~\cite{kananukul2015building}. This same study also suggested that perceived entertainment benefits (e.g., browsing social media for fun or relaxation) did not influence trust in social media platforms~\cite{kananukul2015building}, and perceived entertainment was not a predictor of trust in information on social media. These findings indicate that social media users were unlikely to miscategorize entertaining content with credible and/or trustworthy content on social media. Additionally, Ma et al. found that general perceived social support from a group on social media was associated with trust in that group~\cite{ma2019when}.

\textbf{Risk factors.} Work has explored several nuances that exist between risk factors and trust, including the impact that perceived risks, risk-taking attitudes, and privacy-related risks have on trust. For example, risk-taking attitudes, such as sharing thoughts with group members on social media, were found to increase trust in those Facebook groups~\cite{ma2019when}. Furthermore, encountering perceived risks on a social media platform, such as coming across a seemingly fraudulent account, significantly reduced users' trust in acquiring information from that platform in general~\cite{mohd2020acquiring}. Finally, Malik et al. found that privacy awareness, or the cognizance and understanding of various aspects of privacy on social media, positively influenced users' trust in Facebook~\cite{malik2016impact}. 

\textbf{Homophily.} There is mixed evidence regarding the influence of \textit{homophily}, defined as how similar an individual is to others in the group, on trust in social media users/groups. Specifically, Ma et al.~\cite{ma2019when} found that homophily was not predictive of trust in Facebook groups. Yet, Leonhardt et al.~\cite{leonhardt2020we} suggested that homophily was positively related to trust in other Facebook users. One potential explanation for the discrepancy in results from these two studies may be the differences in measurements.  %Specifically, Ma et al.'s study measured trust via dimensions of care (\textit{``other members of this group care about my well-being''}), reliability (\textit{``other members of this group can be relied upon to do what they say they will do''}), integrity (\textit{``other members of this group are honest''}), and risk-taking (\textit{``I feel comfortable sharing my thoughts in this group.''}). On the other hand, Leonhardt et al.'s study assessed participants' responses to the statements \textit{``I could trust this person (on social media),''}, \textit{``I could believe in this person (on social media),''}, and \textit{``I have confidence in this person'' (on social media)}. 
Major differences exist between the measurements of the two studies, as one study~\cite{ma2019when} assesses \textit{group-based trust} (by asking to what extent participants agree with the statement that ``other members of this group are honest''). In contrast, the other~\cite{leonhardt2020we} assesses \textit{individual-based trust} (by asking if an individual participant has confidence in a particular person on social media). In a different context, Pentina et al.~\cite{pentina2013antecedents} found that Twitter users who believed their personality traits aligned with Twitter's online brand personality were more likely to trust Twitter in general. This is because trust is cultivated among members who believed their own identities aligned with the specific social media brand's identity, according to Pentina et al.~\cite{pentina2013antecedents}. 

\textbf{Social ties.} \textit{Social ties} are the (perceived) strength of social relationships within an individual's social networks~\cite{hsiao2016exploring} and can be categorized into either strong or weak ties~\cite{granovetter1973strength}. Three studies in our corpus suggested that social ties were a significant trust antecedent~\cite{gong2018experienced, gong2018experienced, wang2013trust}, whereas one study in our corpus disagreed~\cite{fogel2019trust}. Wang et al. found that participants' trust towards information on social media is directly correlated with the social ties between the participant and the profile that posted the information~\cite{wang2013trust}. Likewise, Gong et al. found that social ties among members of WeChat played an important role in trust in the WeChat platform in general~\cite{gong2018experienced, gong2021gender}.  

\subsubsection{Trustee-related Antecedents}
Another type of antecedent of trust has to do with the attributes and characteristics of the trustee; an antecedent category we named \textit{trustee-related antecedents}. 
 
\textbf{Transparency.} One trustee-related antecedent of trust revolves around the transparency of the trustee (e.g., information). In this context, transparency is defined as the disclosure of source materials during information production. In general, our findings suggest that transparency is associated with higher trust in a trustee. In a number of studies, transparency has involved presenting information alongside the URL of the original source, which significantly increased people's trust in that information~\cite{baxter2019scottish, chauhan2020trustworthiness, epstein2020will, flintham2018falling, heuer2018trust, wang2013trust}.

% Transparency is broadly defined as ``disclosing information about the journalistic process of news production including decisions, biases, and corrections, should be understood as a means for news organizations to be accountable.'' 

\textbf{Social credibility.} 
The popularity behind a piece of information on social media, often measured by the number of likes, comments, and shares the post has, is also known as \textit{social credibility}. Unfortunately, using source credibility as a way to assess the validity of information online can often lead to the dissemination of unverified information~\cite{tandoc2018defining}. According to a few studies in our corpus, this is because social credibility increases individuals' trust in information found on social media~\cite{barakat2021empirical, wang2013trust}. The authors explained that perceived social credibility decreases users' verification behaviors online, as information shared by others in one's network is more likely to be perceived as credible and, thus, less likely to be personally verified.
 
\textbf{Content \& writing style.} Additionally, the content and writing styles of information online are commonly-examined factors that influence trust. This can include the specific topics that are written about (e.g., informational vs. entertaining information; health vs. political topics)~\cite{choi2017trust, gierth2020attacking, kim2021event, paige2017influence, zhou2021building} and how these topics are introduced and discussed (e.g., wording)~\cite{flintham2018falling, heuer2018trust, kim2021event, shariff2017credibility}. Furthermore, the \textit{level of detail} of the content also influences people's trust in information. For example, in the context of e-commerce, Sari et al.~\cite{sari2018effect} found that Facebook pages containing detailed product information significantly increased users' trust in that page. And yet, another study found that the level of detail of an advertisement on Facebook was \emph{not} directly correlated with participants' trust in the Facebook platform as a whole~\cite{dogruel2019too}. The study found that Facebook advertisements with a medium level of detail led to more trust in the platform than advertisements with a high level of detail (the assessment of levels of detail was determined by participants), especially if the detailed advertisement was highly personalized based on users' online activities (e.g., the websites they visited). %The authors argue that overly-detailed advertisements imply that the platform monitors users' online behaviors (e.g., the websites they visit, geo-location, etc.), likely decreasing users' trust in the platform.

\textbf{Visual \& emotional appeal.} Multi-media and visual components embedded into information online can also increase users' trust in that information. This is because these components often invoke positive emotional sentiments among users, which contributes to increased trust in that information~\cite{chauhan2020trustworthiness, heuer2018trust, kim2021event}. According to the authors, this increase in trust can be attributed to gratification theory, as visually- and emotionally-appealing stimuli better satisfy users' media needs, leading to more trust~\cite{kim2021event}.  

\textbf{Contextual factors.} Contextual factors, including the relevancy of information\cite{chauhan2020trustworthiness}, the timeliness of information~\cite{chauhan2020trustworthiness, chen2016members, huang2020responding}, and the perceived usefulness of information, positively influence trust in information from social media. However, one study focused on COVID-19 information behaviors on social media found that the timeliness of information had no significant effect on information trust~\cite{tayal2021reliability}.

\textbf{Social group attributes.} Our review also shows how group factors can influence people's trust in online groups. For example, Ma et al.~\cite{ma2019when} examined group differences in trust within Facebook groups. Their study found that \textit{group size} was the most significant predictor of trust in Facebook groups, as trust was inversely correlated with the group size. Likewise, \textit{group tenure}, or how long a group has existed and operated on social media, was also shown to predict trust in social media groups. Specifically, the longer the group tenure, the more trust users placed in that group. Finally, the \textit{group category} was shown to be another antecedent of trust, as users often trusted family/friend groups more than interest- or location-based groups.  

\textbf{Platform differences.} Prior work comparing trust across different social media platforms has collectively suggested that people hold different levels of trust in each platform~\cite{cooley2019effect, fogel2009internet, paige2017influence, wang2013trust}. This finding highlights the nuanced differences that exist between platforms that are often overlooked in social media research. These results collectively call into question the notion of generalized trust in social media, as users' trust across platforms can drastically differ. In essence, the distinct features, characteristics, and limitations present among each individual platform need to be more heavily considered in the future, as there is no ``one size fits all'' in trust and social media research. 
 
\begin{summary}
The antecedents of trust in social media in our corpus were grouped into 1) demographic antecedents, 2) trustor-trustee interaction antecedents, and 3) trustee-related antecedents. For the \textbf{demographic antecedents}, age and gender were the most commonly-examined factors. Yet, both of these factors had \textbf{mixed evidence} across papers as to whether or not they were significant trust antecedents. Unlike age and gender, race and ethnicity, geo-location, and political ideology were found to be important antecedents of trust across studies. Very few studies examined education, income, and health literacy (in the context of health-related information) as demographic antecedents. Furthermore, among the \textbf{trustor-trustee interaction antecedents}, social media use was the most prevalent factor examined, followed by prior experiences and attitudes, familiarity with the trustee, and risk factors; fewer studies explored social ties, homophily, and perceived benefits as antecedents of trust in social media. Besides prior (positive) experiences and attitudes, which have been shown to positively influence trust in social media across studies, there is \textbf{mixed evidence} as to whether these trustor-trustee interaction factors are truly trust antecedents. These results indicate that no consensus has yet been achieved regarding what factors (might) influence people's trust in social media, necessitating more research moving forward. Finally, regarding the \textbf{trustee-related antecedents}, factors such as transparency (availability of URLs to information sources), social credibility of information, and visual and emotional appeal embedded into information overall \textbf{positively} influenced trust in social media information. In terms of social group attributes, the smaller the social group and the longer the group tenure, the more trust users placed in the social group. Studies also found that social group category and platform differences influence people's trust in information from social media platforms. Nuanced features and characteristics of the trustee, such as the specific social media platform used, the nature of the social media group, and the characteristics of the individuals sharing the information online, are essential to unpack to better understand trust. 
\end{summary}  

%%%%%%%%%%%%%%%%%%%%%%%%%%%% New Subsection%%%%%%%%%%%%%%%%%%%%%%%%%%%%%%%%%%%%%%%%%%%%%%%%%%%%%%
\subsection{Consequences of Trust}
\label{subsec:impact}

\begin{figure*}[!ht] %[!ht] htbp
  \centering
  \includegraphics[width=0.58\linewidth]{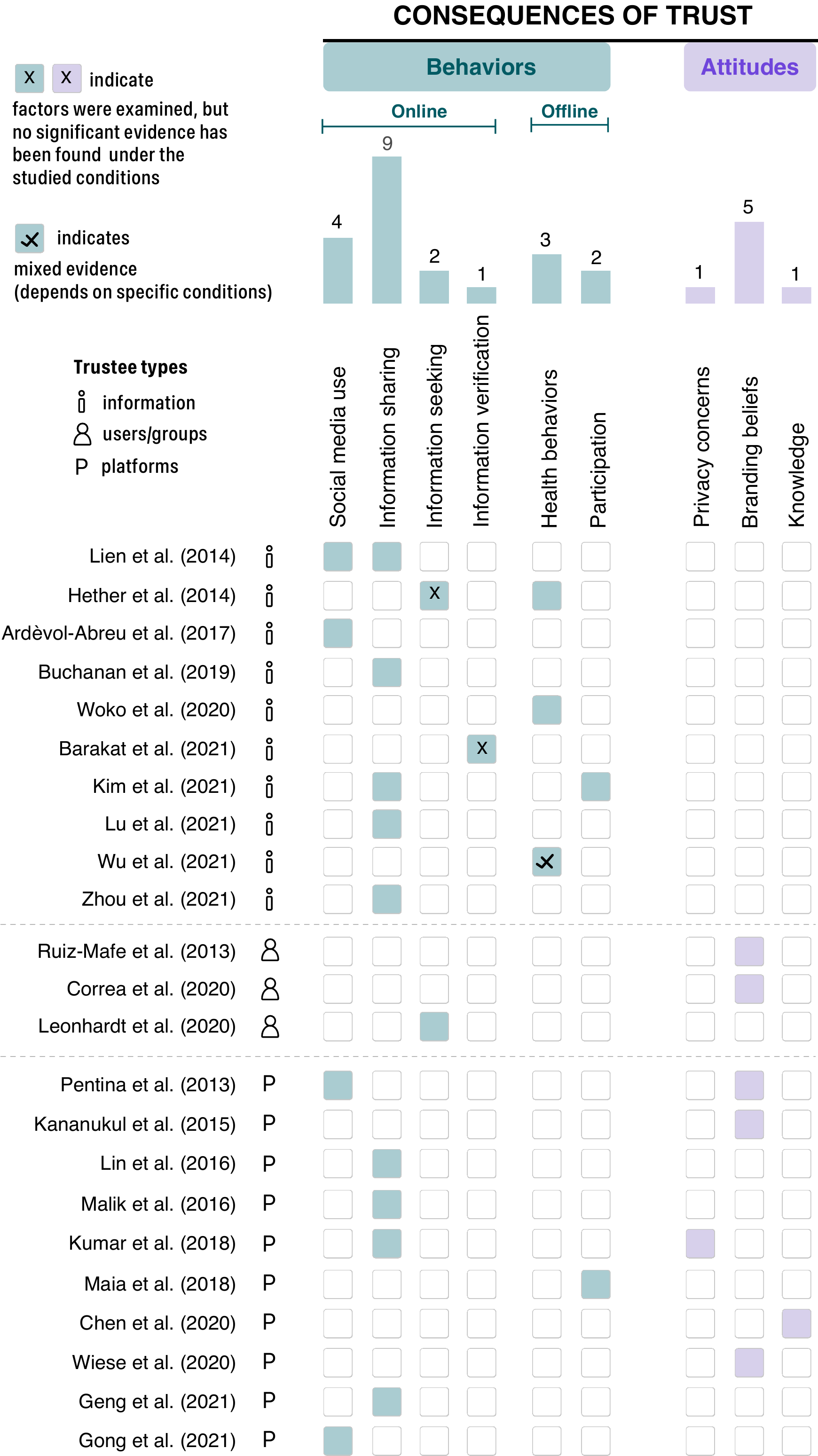}
  \caption{An overview of how trust influences people's online and offline behaviors and attitudes.}  
  \label{fig:consequences}
\end{figure*}

As compared to papers that measured the antecedents of trust, fewer studies (n=23) in our review investigated the consequences of trust in social media. By consequences of trust, we refer to changes in users' attitude or behavior as a result of trusting information found on social media, other social media groups/users, or the social media platforms themselves. These consequences can be categorized into three broad dimensions: online behaviors, offline behaviors, and attitudes spurred by trust, as shown in \autoref{fig:consequences}. Similar to what we mentioned about the findings on trust antecedents, it is important to note that most studies that examined the consequences of trust in social media utilized observational or survey methodology, which ultimately limits any causal conclusions that can be drawn from the research. Nevertheless, based on the theories, hypotheses, and assumptions proposed by each relevant paper, we characterize the consequences associated with trust in social media.

\subsubsection{Online Behaviors}
Online behaviors encompass a wide range of information-technology-mediated interactions and activities. These include \textit{social media use}, which is typically measured by the frequency of using social media for information; \textit{information sharing}, which is characterized by users' likelihood of sharing news and/or posts on the social media platform; \textit{information seeking}, which pertains to the conscious effort to acquire information to bridge a knowledge gap; and \textit{information verification}, which refers to evaluating the trustworthiness of information.  

\textbf{Social media use.} The studies in our corpus show that trust in the information on social media~\cite{ardevol2017effects, lien2014examining}, as well as trust in the social media platforms themselves~\cite{gong2021gender, pentina2013antecedents}, both play a role in increasing people's use of social media. However, as previously mentioned, there is mixed evidence on whether social media use influences trust in information on social media. This suggests that the directionality of this relationship may be important to consider and should be further assessed in future research.

\textbf{Information sharing.} All the studies in our corpus suggest that trust in the information on social media~\cite{lien2014examining, lu2021source, kim2021event, buchanan2019spreading} and trust in the platforms themselves~\cite{geng2021effects, kumar2018interplay, lin2016health, malik2016impact} positively influence people's intention to share information, news, and opinions on social media~\cite{zhou2021building}. 

\textbf{Information seeking.} Hether et al.~\cite{hether2014s} aimed to examine if trust in a social media platform was positively associated with information-seeking behaviors on the platform. However, no significant effect was found. On the other hand, one study found that trust in other social media users positively influenced participants to seek user-generated product information on social media~\cite{leonhardt2020we}.  

\textbf{Information verification.} Verification behaviors include activities such as checking information completeness and comprehensiveness, searching for additional sources to validate the information, considering the biases/agenda of the original authors, and investigating if the information was current and up-to-date, were all unaffected by users' trust in information~\cite{barakat2021empirical}. One study found that trust in information on social media does not affect user verification behaviors, though this trust decreases users' probability of verifying fake news on social media~\cite{barakat2021empirical}. 
 
\subsubsection{Offline Behaviors}
Trust in social media not only influences people's online behaviors but also affects people's offline behaviors. Offline behaviors studied in our corpus include health behaviors (i.e., actions individuals take that affect their health) and participation in events outside of social media. 

\textbf{Health behaviors.} Overall, there is mixed evidence on how trust in information from social media influences people's health behaviors. For example, one study showed that trust in information from WeChat significantly predicted people's level of compliance with health behaviors (e.g., performing preventive measures to mitigate risks of getting coronavirus)~\cite{wu2022exploring}. However, this same study also found that trust in information from Weibo (another social media platform) is negatively associated with people's compliance with health behaviors~\cite{wu2022exploring}. These contradictory findings suggest that people's health behaviors and levels of compliance were dependent on the social media platform that was utilized. Additionally, one study has found that trust in health information on social media is positively associated with users' mental health (e.g., being happier about pregnancy~\cite{hether2014s}). 
 
\textbf{Participation.} Offline participation in events has also been shown to be affected by trust in the information on social media. For example, high levels of perceived trust in event information on social media increases users' willingness to participate in that event offline~\cite{kim2021event}.

\subsubsection{Attitudes}  
In addition, a few studies have examined how trust in social media influences specific attitudes, such as social media users' privacy concerns, branding beliefs, and persuasion knowledge. Specifically, Kumar et al.~\cite{kumar2018interplay} suggested that trust in Facebook significantly impacted privacy concerns about Facebook. Other work found that trust in social media platforms has a positive influence on people's brand trust (defined as the willingness of the average consumer to rely on the ability of a brand to perform its stated function) and on people's online advertising attitudes~\cite{kananukul2015building, pentina2013antecedents, wiese2020determining}. Additionally, two studies found that trust in YouTube content creators~\cite{correa2020influence} and certain fan pages~\cite{ruiz2014key} positively influenced branding beliefs and loyalty. Furthermore, Chen et al. found that trust in Facebook significantly reduced people's persuasion knowledge, or the awareness of and skepticism toward messages and advertisements~\cite{chen2020consumer}. 
 
\begin{summary}
Unlike the mixed results regarding the antecedents of trust in social media, the results gathered from papers that examined the impacts of trust on people's behaviors and attitudes tended to be more consistent. Our findings show that trust in social media significantly \textbf{positively} impacts people's online behaviors (e.g., social media use frequency, intention to share information and news on social media) and offline participation in events. However, trust in social media \textbf{negatively} influences users' information verification behaviors (i.e., the probability of verifying fake news seen on social media). Additionally, there is \textbf{mixed evidence} as to how trust in social media influences people's \textbf{health behaviors}. 
\end{summary}

\subsection{Distrust and Mistrust in Social Media}
\label{subsec:distrust_mistrust}
In our corpus, no paper solely examined distrust or mistrust in social media. A total of 16 papers mentioned both trust and distrust. Among these 16 papers, most of them (n=13, 81\%) viewed trust and distrust as two extremes of the same dimension and did not specifically examine distrust (only briefly mentioned distrust). Only two papers~\cite{cheng2020encountering, heuer2018trust} specifically examined both trust and distrust, and both studies considered trust and distrust as two distinct concepts. Furthermore, we found that one study~\cite{baxter2019scottish} considered mistrust to be the opposite of trust.

Below we examine the two studies that specifically examined trust and distrust. Cheng et al.~\cite{cheng2020encountering} assessed trust in Facebook by measuring \textit{integrity} and \textit{competence}, while also assessing distrust in Facebook by measuring \textit{skepticism} and \textit{vulnerability}. Their study~\cite{cheng2020encountering} shows that trust and distrust in Facebook concurrently and distinctly influenced users' intensity of use. Furthermore, trust had a stronger impact on use intensity of Facebook than distrust. In other words, the degree to which distrust decreased Facebook use intensity was not as significant as the degree to which trust increased use intensity. Another experimental study~\cite{heuer2018trust} focused on ``false trust'' and ``false distrust'' as two kinds of errors in trust. In their study context, \textit{false trust} refers to incorrectly trusting information even though it should not be trusted, whereas \textit{false distrust} refers to incorrectly distrusting information even though the information was credible. Their study explored how high-school students provided trust ratings for online news, focused on high-school students. The authors found that participants' general belief on trusting others can predict their trust rating of a Facebook post, and suggested interventions for those prone to false trust and false distrust.

%%%%%%%%%%%%%%%%%%%%%%%%%%%% New Section %%%%%%%%%%%%%%%%%%%%%%%%%%%%%%%%%%%%%%%%%%%%%%%%%%%%%%
\section{Discussion}
\label{sec:discussion} 

We summarize our findings in \autoref{fig:framework}, which shows an overview of research threads related to trust in social media, as well as recommendations for future work. This figure highlights the three trustees existent in trust and social media research, including \includegraphics[height=2.5mm]{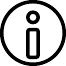} information on social media, \includegraphics[height=2.5mm]{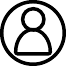} social media users, and \includegraphics[height=2.5mm]{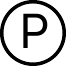} social media platforms. It is crucial for researchers to clearly define who and what the target(s) of trust are and create measurements that assess these trustees accordingly. This figure also emphasizes that future research should clearly define the relationship between trust concepts and the antecedents and consequences of trust. Additionally, {\autoref{fig:framework} conveys the importance of utilizing methods beyond surveys, to enable more nuanced investigations of these relationships. Collectively, {\autoref{fig:framework} and the other diagrams presented in this paper (\autoref{fig:antecedents} and \autoref{fig:consequences}) provide an organization and characterization of key antecedents and consequences of trust in social media that have been studied in prior work. The synthesis of prior work and conceptual mappings in these diagrams can help guide future empirical study design, the creation of trustworthy social media platforms, content, and interventions, and analytic comparisons of future work to prior studies in this area.

\begin{figure*}[h]
  \centering
  \includegraphics[width=0.9\linewidth]{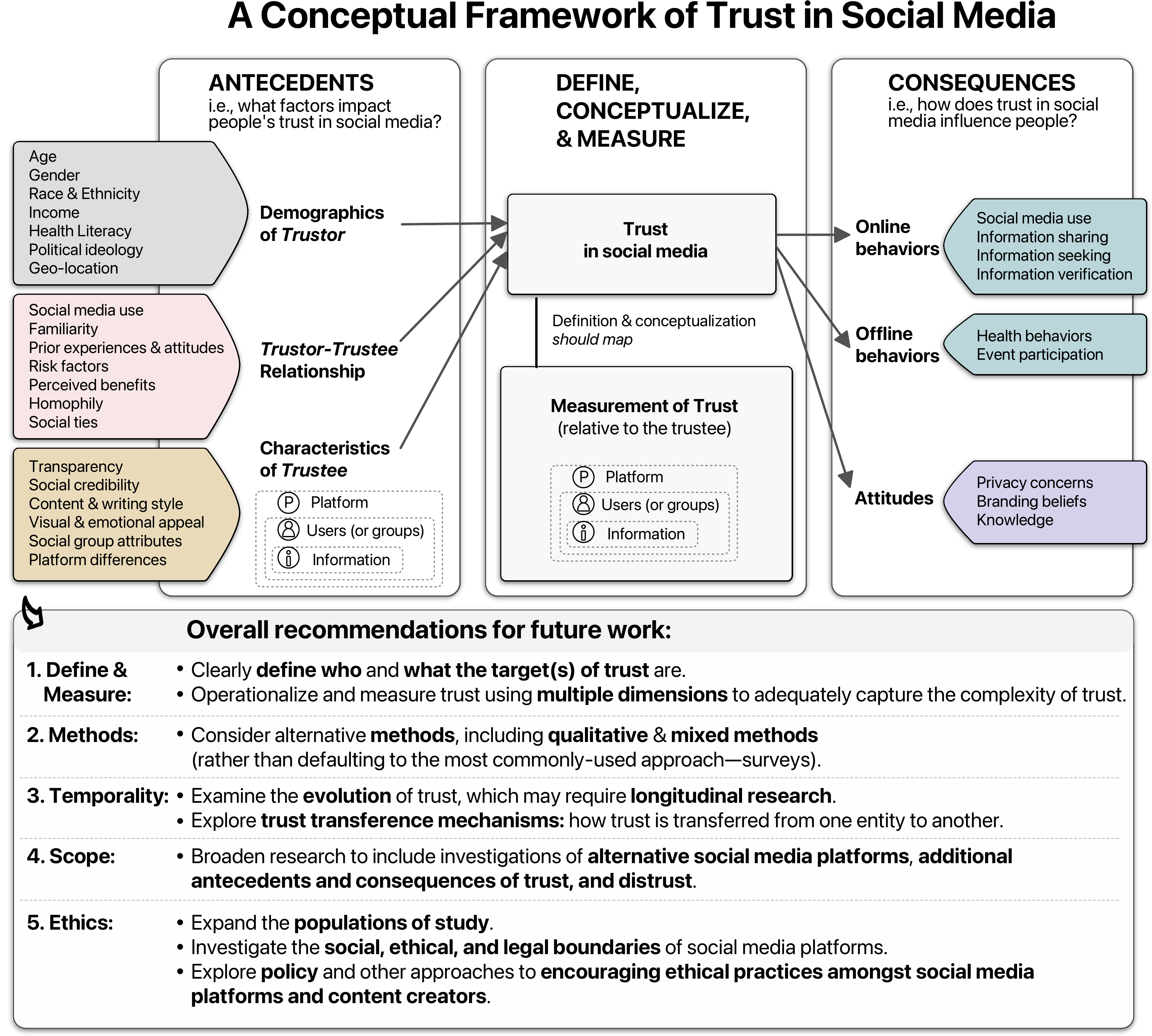}
  \caption{A conceptual framework characterizing key areas of focus when studying trust in social media, and overall recommendations for future work.} 
  \label{fig:framework}
\end{figure*}

Below, we describe in detail the research gaps and future research opportunities that should be explored to 1) concisely conceptualize and measure trust and distrust (\autoref{subsec:discussion-concept}), 2) turn attention to the temporal facets of trust ({\autoref{subsec:discussion-temporal}), 3) expand the scope of what is studied in the context of social media trust ({\autoref{subsec:discussion-scope}), and 4) design future social media information, systems, and interventions (\autoref{subsec:discussion-design-implications}). This discussion can help guide future work in HCI and related fields that seek to deepen our understanding of how and why people trust social media, the consequences of such trust, and implications for the design of social media platforms, content, and interventions. Finally, we conclude by addressing ethical considerations in this field of research ({\autoref{subsec:discussion-ethical}).
% \todo{Added a new diagram Figure 7 and two paragraphs to address R3's comments that ask for a visual to synthesize findings and future directions.}

\subsection{Conceptualizing Trust Concepts}
\label{subsec:discussion-concept}
 
\textbf{Defining and measuring trust in social media.} Our results show that more than half of the surveyed papers did not provide clear definitions of trust and/or did not contextualize the trustee in the context of social media. Unaddressed, we argue that these issues will undermine future research, as ambiguous trust definitions and/or trustees make it challenging to interpret and compare findings across studies. Therefore, future work should clearly specify the context and the trustee in the definition of trust. 

Furthermore, we advocate for future research to ensure that trust definitions align with and guide the trust measurements used in each study, as proper translation of trust into assessment is paramount to ensure construct validity. Our results have shown that trust in social media covers a diverse set of conceptualizations, such as a willingness of the trustor, a feeling, and a decision. Electing to use one conceptualization over another will influence how trust is measured and, consequently, will affect the output of the study. Choosing a proper conceptualization and definition is largely dependent upon the research goals. For example, if conceptualizing trust as a \textit{``willingness to be vulnerable''}, then a measure is needed that assesses a participant's level of ``willingness'' and the nature of that ``vulnerability''. By specifying the precise conceptualization and definition of trust, more pointed research questions arise. For example, following our previous example of trust being defined as a \textit{``willingness to be vulnerable''}, particular questions emerge, such as what characteristics of content lead trustors to develop an openness to being vulnerable? How do these characteristics potentially vary between social media platforms? How is this vulnerability manifested in the trustor? And what are the implications of being vulnerable in this way? These pointed questions invite more nuanced and informative answers and encourage more qualitative methodology outside of just surveys and/or experiments. Prior research has shown the subtleties that can be uncovered using qualitative research when assessing the meaning of trust and how trust is formed in the context of social media~\cite{zhang2022shifting}. Therefore, qualitative research outcomes could be further used to generate measurements, models, and theories for future work.

\textbf{Examining the relationship between trust, distrust, and mistrust in social media.} Our results also show that very few papers explored distrust or mistrust in social media. On top of that, of the papers that did mention both trust and distrust, most considered trust and distrust to be two opposites of one concept. Yet, outside of the field of HCI, many trust scholars have advocated for the notion that trust and distrust should be treated as two distinct concepts, rather than opposites~\cite{cheng2020encountering, harrison2001trust, lewicki1998trust, tang2014distrust}. For example, in the context of political trust, Jennings et al.~\cite{jennings2021trust} differentiate these concepts by describing the diverging attitudes and behaviors that accompany trust versus distrust. Specifically, when people trust a political system, they have accompanying feelings of confidence in, and commitment toward, that system. In contrast, these same researchers explain that distrust comes with a separate set of attitudes that are more than just the opposite of the attitudes that accompany trust~\cite{jennings2021trust}. Rather, distrust in political systems is accompanied by negative emotions, such as contempt, insecurity, fear, anger, alienation, and cynicism~\cite{jennings2021trust}. While trust may inspire loyalty to a system, distrust can foster disengagement from, or even rebellion towards, that same political system. Although researchers have conducted conceptual work to explicate the differing meanings and implications of trust and distrust in political domains, such work has been sparse in the context of social media research. 

Furthermore, in terms of mistrust, our findings show that only one study in our corpus examined mistrust and the study considered mistrust as the opposite of trust~\cite{baxter2019scottish}. However, some trust scholars have argued that mistrust is misplaced trust whereby a trustee ``defaults'' or betrays a person's trust (and mistrust does not imply the opposite of trust)~\cite{marsh2005trust}. Therefore, more research is needed to investigate the relationship between trust, distrust, and mistrust in social media, the dimensions of each, the nuances of the ways in which people's trust, distrust, and mistrust are formed, as well as the attitudes and behaviors that these \textit{trust concepts} in social media invoke. Ignoring distrust and/or mistrust in social media may yield incomplete estimates regarding the consequences of trust~\cite{tang2014distrust}.

\subsection{Attending to Temporal Facets of Trust} 
\label{subsec:discussion-temporal}

\textbf{Shifting attention to the evolution of trust in social media.} Our results show that almost all studies in our corpus used study designs involving data collection at a single time point. In contrast, only three studies conducted the two-wave measurement, and no studies in our corpus have adopted a longitudinal approach to understanding the temporal facets of trust in the context of social media. Note that two waves of measurement are not considered longitudinal studies~\cite{ployhart2010longitudinal,ployhart2014two}. However, given that trust in the information found on social media, users, and platforms does not necessarily remain static and instead changes and develops over time~\cite{zhang2022shifting}, it is essential to evaluate the \emph{evolution} of a relationship between the trustor and trustee and the long-term effects of trust. As our findings have shown, trust can impact the use of social media and could, in turn, lead to a change in trust antecedents, such as a change in social media use. This reciprocal relationship highlights that trust can evolve over time, shaping users' behaviors and their perception of social media. Such investigations are necessary to build a more comprehensive understanding of the development, breakdowns, and implications of trust in the context of social media.  

\textbf{Exploring trust transference mechanisms.} In the context of social media where multiple trustees exist, \textit{trust can be transferred}, meaning that an individual's trust in one entity (e.g., social media content) can be derived from their trust in a related entity (e.g., a social media platform)~\cite{stewart2003trust}. Trust transference has been explored in other contexts, such as in business and online banking~\cite{handarkho2020understanding, stewart2003trust}. However, besides one study conducted by Pentina et al.~\cite{pentina2013antecedents}, trust transference mechanisms were rarely examined in our corpus. Specifically, Pentina et al.~\cite{pentina2013antecedents} found that those who had a high level of trust in Twitter as a platform often transferred this trust to the brands they followed on Twitter. In this case, trust transfers from the social media platform to other parties on the platform. In short, the trust transference mechanisms have not been well studied in the context of social media and, thus, require additional research. For example, future work may explore factors that drive these transference mechanisms which will help us better understand how trust is transferred. Understanding the mechanisms and antecedents of trust transference in social media will help researchers develop strategies to promote or impede trust transference.  

Furthermore, future work should focus on the nuances of trust transference mechanisms to accurately specify the trustee (i.e., information, users, platform) being referred to. Likewise, special care should be taken to ensure that trust in each trustee is measured independently, rather than trying to combine and interpret trust across multiple trustees in one composite variable, as we highlighted in \autoref{subsec:measurements}.

\subsection{Broadening the Scope of Social Media Trust Research} 
\label{subsec:discussion-scope} 

\textbf{Expanding the social media platforms examined.} Our results suggest that many studies have operationalized trust in social media as a general concept without specifying distinct social media platforms (also see in \autoref{tab:platform}). However, it is important to note that platform differences shape people's trust in social media, as our results suggested. That is, social media users hold varying degrees of trust across social media platforms. Given these differences in trust, future work may consider comparing and contrasting users' trust across platforms to better understand the mechanisms through which trust is formed on social media. Such work is also needed to identify platform-specific opportunities that could be incorporated to enhance users' trust. Additionally, our findings showed that Facebook and Twitter were the most frequently-studied platforms by far, opening the door for future work to examine other emerging platforms.

\textbf{Expanding the understanding of antecedents of and consequences of trust and distrust.} Our findings show that existing work has examined a variety of trust antecedents and consequences of trust in social media. Nevertheless, conflicting results exist regarding a few trust antecedents, namely social media use and familiarity. Specifically, when examining the antecedents of trust, some studies suggested that higher usage of social media increases trust in the information found on social media, whereas other studies either found the exact opposite relationship or no significant relationship at all. And yet, multiple studies that examined the consequences of trust converged on the finding that a higher level of trust in either information on social media or the platforms themselves increased people's social media use. We speculate that trustors' experiences and content seen on social media may play important roles in understanding these conflicting results. For instance, if users have seen a large amount of misinformation on social media, they could lose trust in that particular platform~\cite{zhang2022shifting}. Future work is needed to clarify these mixed findings and identify additional factors that help explain them.
 
Reflecting on our results on the consequence of trust, we can see emerging evidence supporting the idea that trust in social media can influence users' online behaviors, offline behaviors, and attitudes. Still, there are fewer studies on the consequence of trust than those examining trust antecedents. Particularly, only a few studies in our review explored how people's trust in the information found on social media influences their information-seeking and information-verification behaviors. As such, more work is needed to examine the extent to which trust shapes people's beliefs and interpretations of information on social media. Given that the consequences of trust impact users' everyday attitudes and behaviors, important opportunities for future research remain, as expanding our understanding of the consequences of trust in social media may help us better understand and predict users' behaviors and attitudes.

\subsection{Design Implications} 
\label{subsec:discussion-design-implications}

Based on our findings, we provide several implications for future work that seeks to design trustworthy social media systems, interventions, or social media posts.

First, \textbf{consider how trustor characteristics impact trust formation:} Our review  summarizes a set of personal factors that influence people's trust towards the information found on social media, social media users and groups, and the platforms themselves. These findings suggest the need to take the individual differences between social media users into consideration when designing and delivering information and/or interventions. For example, Hispanic communities tend to trust doctors and community leaders of the same ethnic background, but they are less likely than White populations to trust the government and social media influencers in the context of COVID-19~\cite{kricorian2021covid}. These trends suggest that culturally-competent approaches~\cite{zakaria2003designing}, such as enlisting community-trusted Hispanic doctors and community leaders to develop and disseminate public health information, may be much more effective than current, more general approaches. Future research may consider exploring these culturally-competent approaches to ensure that they are relevant and effective for specific populations and workplaces. In addition, an emerging body of work has begun to explore the use of automated conversational agents and language models to identify unique traits in social media users' communication styles~\cite{schwartz2013personality}. Indeed, communication styles differ among people of different backgrounds and cultures, meaning general-purpose approaches to designing and delivering information and/or interventions may have adverse effects or fail to feel relevant to certain populations. To this end, future work should further explore the nuances in the interplay between users' backgrounds and communication styles and how they could affect what people trust online. 
 
Second, \textbf{consider how trustee characteristics and broader contextual factors intersect to build trust:} Our review shows that a wide range of factors related to the trustee positively influences trust in social media information, including transparency (e.g., URLs to information sources) and social credibility of information (measured by the number of likes, comments, and shares). These factors also varied across information-related, user-related, and platform-related attributes. As such, building trust with users is a complex task that requires effort to ensure that these attributes are fully realized. Accordingly, future design solutions that address trust issues in social media should consider various possible trustee scenarios. By ``trustee scenarios'', we refer to examining the aforementioned attributes (e.g., transparency, social credibility, contextual factors, and so forth, as listed in \autoref{fig:antecedents}) in concert, and with respect to different trustees (e.g., information, other users, and/or the platforms themselves). For example, a multiple-feature-based approach that provides indications of the reputation of the information source (and the source of the source, if any), the total number of likes, comments, and shares that a post has, the sharer's reputation, and the styles of language and writing may help provide a comprehensive view. Another possible approach is to utilize third-party entities and crowdsourcing measures to monitor how social media platforms recommend information that could influence their trustworthiness. For example, recent research has suggested that utilizing crowdsourced data (e.g., general public rating of information outlets) to identify misinformation-producing sources and imputing this data into social media ranking algorithms that rank content within a platform has promise for reducing the amount of misinformation on social media platforms~\cite{epstein2020will}. 

Additionally, our review has also shown that although timely information positively influences people's trust in information from social media, this timeliness of a social media post had no significant effect on people's trust in that post during COVID-19~\cite{tayal2021reliability}. Likely, the high-level uncertainty during a public health crisis further complicates users' credibility assessments and trust formation. All these findings suggest a need for future work that compares and contrasts the contexts in which trust formation occurs to achieve a better understanding of how solutions may potentially be transferred to other contexts. 

In summary, future design that seeks to foster people's trust should consider multiple antecedents of trust in relation to the three different trustees (e.g., information, users, and platform), as well as the socio-cultural-political contexts in which they are embedded.

\subsection{Ethical Implications} 
\label{subsec:discussion-ethical} 
Our work provides insights into the factors that promote trust in information on social media, other users on the platform, and the social media platforms themselves. Ethical considerations are needed to ensure that studies of trust are used to empower users instead of manipulating them. For example, our results showed that integrating more visuals (e.g., graphs, visualizations, elegant design) can increase the perceived trustworthiness of information online. Unfortunately, these strategies may also be used for more nefarious purposes, such as inflating the perceived credibility of misinformation and disinformation online~\cite{villasenor2020people}. Future work should investigate, for example, how misleading visualizations are created to serve what purposes~\cite{zhang2021mapping, zhang2023visualization}, and how charts are used and misused to support an argument someone is making. Potential solutions may include the detection of misleading graphs using machine learning techniques, which is an understudied area. Below, we describe additional ethical implications that our work raises. 

First, it is essential to \textbf{equitably and respectfully engage with marginalized populations.} 
Very few studies in our corpus included populations who disproportionately experience barriers to wellbeing, such as lower-socioeconomic status groups. And yet, many demographic factors, including race, ethnicity, and income, play an essential role in shaping what and whom people trust within social media, as we detailed in our study. Excluding these groups from study samples is fundamentally unethical and can lead to biased findings and, subsequently, the development of systems with embedded bias~\cite{zhang2020understanding}. Therefore, these findings emphasize a genuine need for future work to focus on marginalized communities when studying the formation and implications of trust in social media. However, it is crucial to ensure that members of these communities desire to collaborate with researchers before conducting research or introducing interventions into their communities~\cite{brown2016five}. After all, subjugating these communities to unwanted research is detrimental to their well-being and autonomy. Likewise, in cases where technologies deployed to these communities are taken away from them at the end of the study, distrust in researchers and hesitation to participate in future research is likely to occur among these participants~\cite{resnik2009clinical}. As such, investigation into the long-term and lasting effects of HCI research on these populations is needed.  

Second, there is a need for \textbf{educational interventions that promote ethics among social media users.} When navigating the social media space, social media users should not take the responsibility to be ethical lightly. For example, as our review indicates that when tagging, commenting, and posting on social media, social media users become part of the web's collective intelligence{~\cite{rheingold2012crap}. Consequently, many people have been negatively impacted by other social media users who post, re-share, or comment on false content being presented as facts~\cite{rheingold2012crap}. These issues suggest the need for educational interventions that highlight the importance of ethics in social media and technology use. These interventions may include raising privacy and security awareness regarding how an individual's social media data can be used by the platform, third parties, and other social media contacts~\cite{acquisti2017nudges}; encouraging social media users to verify sources of information before (re-)posting online~\cite{van2022misinformation}; and ensuring that communication maintains the dignity and respect of involved users~\cite{bowen2013using}.

Third, critically assessing \textbf{social media platform accountability} is important. Our findings demonstrate how users' trust in social media platforms is influenced by the credibility and quality of information on social media; this trust (or lack thereof) consequently influences users' future engagement with those platforms. However, social media companies have been criticized for failing to combat the spread of misinformation (e.g., failing to label and remove misinformation) even though they have promised to take action to address these issues~\cite{ISD_failure}. For example, effective November 23, 2022, Twitter is no longer enforcing its policy focused on combating COVID-19 misinformation~\cite{Twitter_noLabelMisinfo}. This is concerning given the extensive repercussions COVID-19 misinformation has had on users' attitudes and behaviors~\cite{lee2022misinformation}. By tolerating the spread of misinformation and disinformation on their platforms, social media companies are failing to protect their users. As such, more research is needed to investigate social media's accountability in addressing issues of misinformation. 
 
\subsection{Limitations}  
One limitation of this work is that our database search was scoped to include six databases. Although measures were taken to ensure that these databases contained a majority of the studies available that focus on trust in social media, it is possible that some studies were not included in the selected databases. Additionally, our corpus is restricted to English-language papers. Therefore, future work may expand the scope of literature review written in other languages.

\section{Conclusion}
Through our systematic review of 70 papers, this work mapped the landscape of literature focused on \textit{trust concepts} (that include trust, distrust, and mistrust) in social media. We identified patterns in this body of work, including trends in study designs used, definitions, conceptualizations and measurements, and the antecedents and consequences of \textit{trust concepts} in social media. This paper calls attention to the complexities and discrepancies that exist within this research field and identifies ways in which future research can tackle these major issues. Our recommendations will help to increase the validity of trust research moving forward as researchers begin to clearly establish what they truly \emph{mean} when they talk about trust in social media. The formation, evolution, and deterioration of trust in the context of social media will continue to be a challenging but crucial research topic and is one that deserves further study.

\begin{acks}
This material is based on work that is funded by an unrestricted gift from Google. We thank the Wellness Technology lab at Georgia Tech for feedback and our anonymous reviewers for their reviews.  
\end{acks}

\bibliographystyle{ACM-Reference-Format}
\bibliography{ref}

% \appendix

\end{document}